\documentclass[useAMS,usenatbib]{mnras}
\usepackage{graphicx}
\usepackage{amsmath}
\usepackage{times}
\usepackage{geometry}
\usepackage{pdflscape}
\usepackage[T1]{fontenc}

\def \aj {AJ}
\def \mnras {MNRAS}
\def \pasp {PASP}
\def \apj {ApJ}
\def \apjs {ApJS}
\def \apjl {ApJL}
\def \aap {A\&A}

\def \nat {Nature}
\def \araa {ARAA}

\newcommand{\kms} {$\mathrm{ km \; s^{-1}}\,$}
\newcommand{\msol} {M$_{\odot}$}

\def\lesssim{\mathrel{\hbox{\rlap{\hbox{\lower4pt\hbox{$\sim$}}}\hbox{$<$}}}}
\def\gtrsim{\mathrel{\hbox{\rlap{\hbox{\lower4pt\hbox{$\sim$}}}\hbox{$>$}}}}
\def\halpha {$\mathrm{H\alpha}$}
\def\ang {$\mathrm{\AA}\,$}
\def\hbeta {$\mathrm{H\beta}$}

\title[Spectropolarimetry of the 2012 outburst of SN 2009ip]{Spectropolarimetry of the 2012 outburst of SN 2009ip: a bi-polar explosion in a dense, disk-like CSM  \thanks{Based on observations made with ESO Telescopes at the Paranal Observatory, under programme 290.D-5006.}}
\author[Reilly et al.]{Emma Reilly$^{1}$\thanks{Email: ereilly528@qub.ac.uk}, Justyn R. Maund$^{2,3}$, Dietrich Baade$^{4}$, J. Craig Wheeler$^{5}$,  Peter H{\"o}flich$^{6}$, \newauthor{Jason Spyromilio$^{4}$, Ferdinando Patat$^{4}$}, Lifan Wang$^{7}$\\
$^{1}\,$Astrophysics Research Centre, School of Mathematics and Physics, Queen's University Belfast, Belfast BT7 1NN, UK.\\
$^{2}\,$Department of Physics and Astronomy, The University of Sheffield, Hicks Building, Hounsfield Road, Sheffield, S3 7RH, UK\\
$^{3}\,$Royal Society Research Fellow\\
$^{4}\,$ESO - European Organisation for Astronomical Research in the Southern Hemisphere, Karl-Schwarzschild-Str. 2, 85748 Garching\\ b. M{\"u}nchen, Germany\\
$^{5}\,$Department of Astronomy and McDonald Observatory, The University of Texas, 1 University Station C1402, Austin, Texas 78712-\\ 0259, U.S.A.\\
$^{6}\,$Department of Physics, Florida State University, Tallahassee, Florida 32306-4350, U.S.A.\\
$^{7}\,$Department of Physics, Texas A\&M University, College Station, Texas 77843-4242, U.S.A.\\
}

\date{Accepted XXX. Received YYY; in original form ZZZ}

\pubyear{2017}

\begin{document}
\label{firstpage}
\pagerange{\pageref{firstpage}--\pageref{lastpage}}
\maketitle

\begin{abstract}
We present a sequence of eight spectropolarimetric observations monitoring the geometric evolution of the late phase of the major 2012 outburst of SN 2009ip. These were acquired with the FORS2 polarimeter mounted on ESO VLT. The continuum was polarised at 0.3-0.8\% throughout the observations, showing that the photosphere deviated substantially from spherical symmetry by 10-15\%. Significant line polarisation is detected for both hydrogen and helium at high velocities. The similarity in the polarised signal between these elements indicates that they form in the same location in the ejecta. The line polarisation ($p\sim$1-1.5\%) at low velocities revealed the presence of a highly-aspherical hydrogen and helium rich circumstellar medium (CSM). Monte Carlo simulations of the observed polarimetry were performed in an effort to constrain the shape of the CSM. The simulations imply that the polarimetry can be understood within the framework of a disk-like CSM inclined by 14$\pm$2$^{\circ}$ out of the line of sight, obscuring the photosphere only at certain epochs. The varying temporal evolution of polarisation at high and low velocities indicated that the fast-moving ejecta expanded with a preferred direction orthogonal to that of the CSM. 
\end{abstract}

\section{Introduction}
\label{sec:intro}
There is a growing body of evidence that suggests that the supernova explosions at the end of the life of massive stars are not spherically symmetric events. The most notable evidence was the resolved asymmetric structure of SN 1987A, as observed in late time HST imaging \citep{Jan2001, Lar2013, Sin2013}. Complex aspherical morphologies have also been observed in Galactic supernova remnants (SNRs), including that of Cassiopeia A \citep{Fes2001,Whe2008, Mil2013, Rest2011b}. Light echoes of SNe \citep{Cro2008, Res2011,Sin2013} and nebular phase spectroscopy \citep{Mae2008} have also been utilised to reveal departures from spherical symmetry. One of the most substantive indicators for asymmetry in the last decade has come from spectropolarimetry of extragalactic SNe. Polarimetry has shown that significant departures from spherical symmetry are present in all types of core-collapse SNe (CCSNe) \citep{Wan2008}. The observations showed that the polarisation often increases with time as the core of the explosion becomes increasingly exposed \citep{Wan1996, Leo2006, Mau2007b, Mau2009, Chor2010}. This suggests that the explosion itself is asymmetric in nature.

In addition to being a powerful probe of the 3D structure of SNe ejecta, spectropolarimetry can also divulge information on the medium surrounding the progenitor stars. This is particularly relevant for the interaction-dominated Type IIn SNe. The spectra of Type IIn SNe are hydrogen rich and characterised by narrow ($\sim$ 1000 \kms) emission lines, and are an indicator that the ejecta are strongly interacting with dense interstellar or circumstellar matter \citep{Sch1990, Fil1997}. They are thought to originate from the deaths of very massive stars that have lost a considerable amount of their envelope through episodic mass-loss \citep[][see also \citealt{Smith2014ARAA}]{Gal2007, Tad2013}. There are, however, other channels that have been proposed (such as through thermonuclear explosions in a dense circumstellar medium; \citealt{Tad2012}).

One of the most controversial and best studied objects of this class is SN 2009ip. It was discovered by \citet{Maz2009} in the spiral galaxy NGC 7259 in August 2009 and designated a supernova before it was re-classified as the extreme brightening of a Luminous Blue Variable (LBV) \citep{Ber2009ATel, Mil2009ATel}. Serendipitous pre-explosion Hubble Space Telescope images dating from 1999 allowed \citet{Smi2010} and \citet{Fol2011} to estimate an initial mass of 50-60 \msol. SN 2009ip then endured three more years of outbursts culminating in two spectacular eruptions in 2012 \citep{Pas2013}.

In July-August 2012, SN 2009ip began to re-brighten once again. This occurred in two phases, the first outburst beginning on 24$^{\mathrm{th}}$ July 2012 (2012a) reached a peak absolute magnitude of $M_{V} \approx-$14 mags. A second re-brightening event began on 23$^{\mathrm{rd}}$ September 2012 (2012b) and took the peak magnitude to $M_{R}=-$18.5 mags, similar to some CCSNe. This outburst was accompanied by evidence of material travelling at speeds of up to 15000 \kms and a spectrum that was reminiscent of the early phases of Type IIn SNe \citep{Mauer2013,Pas2013}. This led many to believe that SN 2009ip had finally undergone a terminal explosion. The high SN-like velocities, however, had previously been observed in 2011 \citep{Pas2013} and the luminosity of the 2012a event was somewhat low for a SN explosion. These observations cast doubt on the supernova explosion scenario. The curious case of SN 2009ip has now been joined by a handful of similar objects, including SN 2015bh \citep{Ofe2016, Sok2016, Eli2016, Tho2016}, SN Hunt 248 \citep{Kan2015} and LSQ13zm \citep{Tar2016}.

The final fate of SN 2009ip has been much debated in the literature. The proposed explanations for the 2012 outbursts fall into 3 scenarios: a terminal CCSN in either the 2012a or 2012b outbursts \citep{Mauer2013, SMP14, Mauer2014} or another ``impostor" event from the eruptions of an LBV star \citep{Pas2013, Fra2013, Mar2014}. An alternative possibility is that the 2012 events were the result of close encounters and the possible merger of an LBV star and a more compact companion \citep{Kas2013, Sok2013}.

Despite the detailed studies of this object, there is still no clear consensus on whether the progenitor star survived the events of 2012. As of yet there has been no evidence of nucleosynthesis products in late-time spectra which would confirm core collapse \citep{Fra2013, Mar2014, Fra2015}. Most recently \citet{Tho2015} have reported that the luminosity of SN 2009ip has now declined below that of the possible progenitor observed in 1999. This possibly favours the terminal SN explosion scenario, or suggests that the star was in outburst during those pre-explosion observations. 

\citet{Mauer2014} have previously published polarimetric observations of this outburst. The large continuum polarisation they observe at the peak of the 2012b event suggests a highly-aspherical interaction zone, from which they infer a disk-like or toroidal circumstellar medium (CSM). Additionally, the ejecta are orthogonal to the plane of the CSM. They propose that the small angular coverage of the CSM requires a large SN-like impact velocity to produce the observed luminosity.

In this paper we present spectropolarimetric observations of the decline of the 2012b event. Section \ref{sec:obs} contains information on the observations and data reduction and in Section \ref{sec:specev} we briefly discuss the spectral evolution. The polarimetric results are then analysed in Section \ref{sec:analysis}. In Section \ref{sec:discussion} we present a Monte Carlo simulation of the geometry of the CSM and present our conclusions in Section \ref{sec:09ip_conclusions}.

\section{Observations and Data Reduction}
\label{sec:obs}
\begin{table*}

  \caption{Table of VLT FORS2 observations for SN 2009ip}
     \begin{tabular}[alignment]{c c c c c c c}
  \hline
  \hline
  Object & Date & Grism & Exposure & Mean Airmass & Epoch\dag& S/N \\
   & (UT) & & (s) & & (days) &\\
  \hline
 $\star$ SN2009IP & 2012 11 7.02 & 300V & $4\times330$ & 1.01 &$+$35 & 550 \\
 $\star$ SN2009IP & 2012 11 7.03 & 1200R & $4\times470$ & 1.04 &$+$35 & 400\\
   LTT1020\ddag  & 2012 11 7.06 & 300V & 10 & 1.12 &  & \\ 
  LTT1020\ddag  & 2012 11 7.06 & 1200R & 30 & 1.12 &  & \\  
  \\
   SN2009IP & 2012 11 14.07 & 300V & $4\times310$ & 1.17 &$+$42 & 375\\
  SN2009IP & 2012 11 14.08 & 1200R & $4\times500$ & 1.26 &$+$42 & 300\\
  LTT1020\ddag  & 2012 11 14.18 & 300V & 10 & 1.06 &  & \\  
  LTT1020\ddag  & 2012 11 14.19 & 1200R & 25 & 1.06 &  & \\  
  \\
 $\star$ SN2009IP & 2012 12 6.04 & 300V & $4\times895$ & 1.43 &$+$64 & 300\\
   LTT1020\ddag  & 2012 12 6.09 & 300V & 15 & 1.01 &  & \\  
   \\
 $\star$ SN2009IP & 2012 12 10.04 & 1200R & $4\times890$ & 1.53 &$+$68 & 200\\
   LTT1020\ddag  & 2012 12 10.09 & 1200R & 40 & 1.02 &  & \\  
   \\
  SN2009IP & 2012 12 15.04 & 300V & $4\times895$ & 1.73 &$+$73 & 150\\
   LTT1020\ddag  & 2012 12 15.09 & 300V & 15 & 1.05 &  & \\  
   \\
SN2009IP & 2012 12 25.04 & 1200R & $4\times890$ & 2.10 &$+$83 &  50\\
   LTT1020\ddag  & 2012 12 25.09 & 1200R & 40 & 1.10 &  & \\  
\hline
\hline
  \end{tabular}
  \vspace{2mm}
  
  \begin{flushleft}
  \dag Relative to UV maximum of the 2012b event which occurred on the 3$^{\mathrm{rd}}$ October 2012 \citep{Mar2014} \\
  \ddag Flux standard\\
  $\star$ Data that were previously published in \citet{Mauer2014}
  \end{flushleft}
  \label{obstable}
\end{table*}

Spectropolarimetric observations of SN 2009ip were commenced  on 7 November 2012, 35 days after the UV maximum of the 2012b event on 3$^{\mathrm{rd}}$ October 2012 \citep{Mar2014}. An outline of the observations is given in Table \ref{obstable}. In total eight observations were conducted over six different epochs. A portion of these data, four spectra over three epochs, have previously been published in \citet{Mauer2014}. These are marked in Table \ref{obstable}. Further observations are presented here, for those data previously published a different approach to the interstellar polarisation is taken. The spectropolarimetric data were acquired using the European Southern Observatory VLT Antu telescope with the Focal Reducer and Low Dispersion Spectrograph (FORS2) instrument in the PMOS mode \citep{FORS}. In PMOS mode, FORS2 operates as a dual-beam optical spectropolarimeter. FORS2 measures linear polarisation with a ``super achromatic" half wave retarder plate that rotates the plane of polarisation before a Wollaston prism separates the light into two orthogonally-polarised beams. Observations were conducted with the retarder plate at four angles: 0.0$^{\circ}$, 45.0$^{\circ}$, 22.5$^{\circ}$ and 67.5$^{\circ}$. The ``striped" PMOS slit mask was used with slit width of 1 arcsec. The 300V and 1200R grisms were used in order to obtain a dispersion of $\sim$3.23 \ang pixel$^{-1}$ and a spectral resolution of $\sim$12 \ang across the spectrum with the former, whilst the latter provided a dispersion of $\sim$0.73 \ang pixel$^{-1}$ and a spectral resolution of $\sim$2.7 \ang. To prevent second order contamination at longer wavelengths an order separation filter (GG435) was used, yielding wavelength ranges of 4450-8650 \ang for the 300V and 5750-7310 \ang for the 1200R grism. 

The data were reduced in IRAF  \footnote{ IRAF is distributed by the National Optical Astronomy Observatories, which are operated by the Association of Universities for Research in Astronomy, Inc., under cooperative agreement with the National Science Foundation -http://iraf.noao.edu/.} using the method outlined in \citet{Mau2007b} and the normalised Stokes parameters ($q$ and $u$) were calculated following \citet{Pat2006}.  In order to increase the signal-to-noise ratio, the spectra were binned to 15 \ang and 5 \ang for the 300V and 1200R spectra, respectively. This was performed prior to the calculation of the Stokes parameters.

The data were corrected for the wavelength dependent chromatic zero-angle offset of the retarder plate, and the polarised spectra were debiased following the prescription of \citet{Qui2012}. At each epoch, calibrated flux spectra were produced using the spectra observed at all four retarder plate angles. The combined spectrum was flux calibrated using observations of a flux standard star, observed with the polarimetry optics in place and with the retarder plate at 0$^{\circ}$.


\section{Spectral Evolution}
\label{sec:specev}
\begin{figure*}
\includegraphics[scale=0.43]{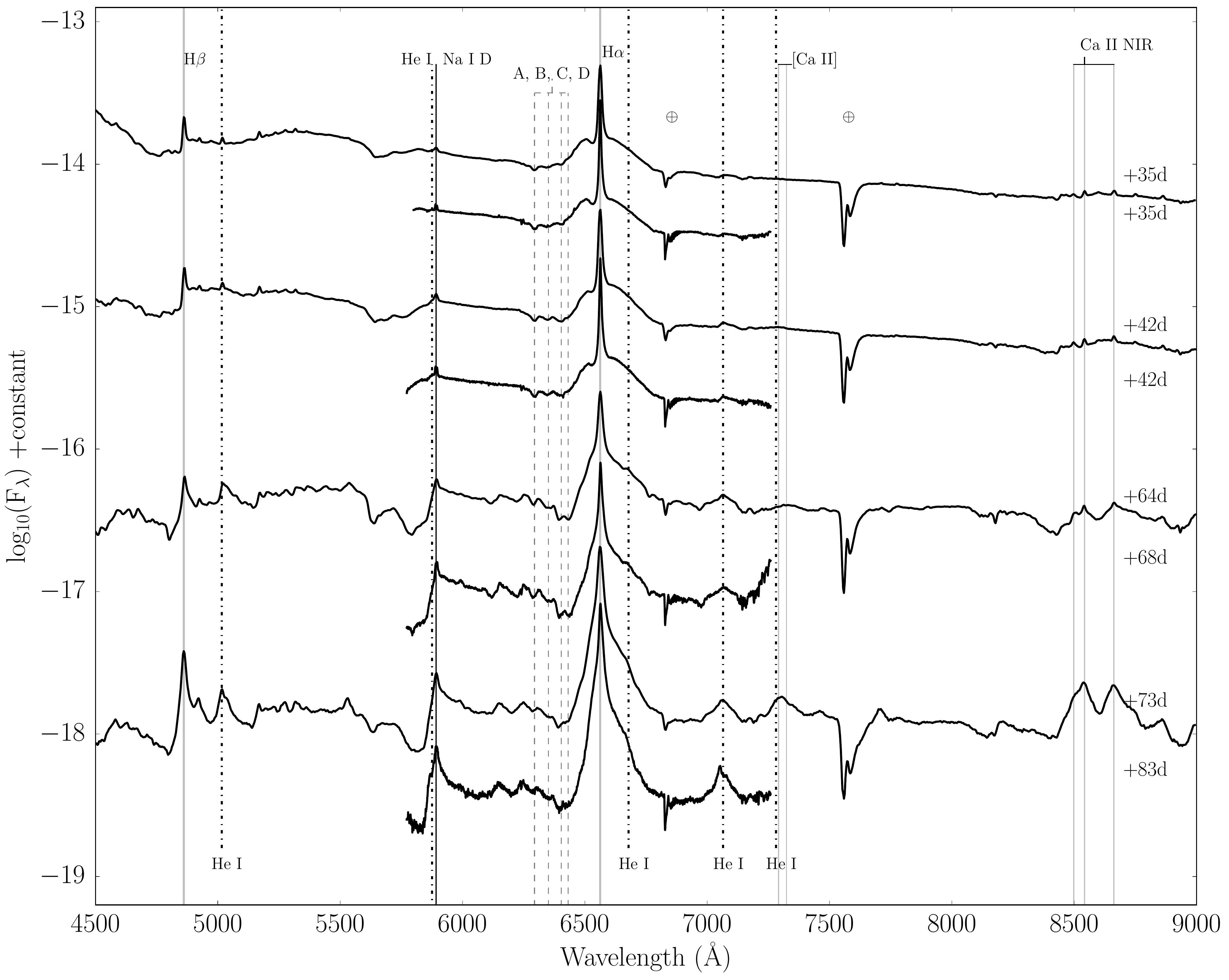}
\caption{Flux spectra of SN 2009ip for the observations between 7 November 2012 ($+35$ days) and 25 December 2012 ($+83$ days) taken with the FORS2 instrument with the $300V$ and $1200R$ grisms. Line identifications indicated are taken from \citet{Fra2013} and \citet{Mar2014}, the vertical lines show the position of the rest wavelength for these lines. Also identified are the high-velocity narrow absorption ``notches" of \halpha: A, B, C and D. Telluric features are indicated by $\oplus$.}
\label{lineid}
\end{figure*}

The flux spectra for the observations taken with the FORS2 instrument between the 7 November 2012 ($+$35 days) and 25 December 2012 ($+$83 days) are presented in Figure \ref{lineid}. Line identifications indicated in the figure are taken from \citet{Fra2013} and \citet{Mar2014}, where the spectroscopic evolution of SN 2009ip from this period can be found.

At $+$35 and $+$42 days \halpha\,\,displays a distinct and asymmetric broad emission (FWHM$\approx$10,200 \kms) with a narrow emission (FWHM$\approx$300 \kms) superimposed. A narrow absorption component can be seen with a minimum at $\sim$$-$1500 \kms at $+$35 and $+$42 days. The absorption decreases in strength until it has disappeared by $+$64 days. At this stage the shape of the broad emission has changed, becoming more reminiscent of SNe in the nebular phase. A broad absorption component with FWHM $\sim$ 10$^{4}$ \kms is apparent at all epochs. Superimposed on this broad absorption are further narrow absorption notches at $-12,300$, $-9700$, $-7300$, $-6000$ \kms, which we label A, B, C and D respectively. In the early epochs the absorption notches A, B and C are strongest, at $+$64 days components C and D increase in strength before notch D becomes weaker again at $+$73 days. These narrow components are more clearly visible in the 1200R observations. \hbeta\, shows a broad absorption trough with a similar multicomponent velocity structure embedded, this is most prominent at $+$42 days. In contrast, \hbeta\,\,does not show the same broad emission that \halpha\,\,does.

A broad blend of He\,{\sc i} $\lambda5876$ and Na\,{\sc i}\,{\sc d} in absorption is observed in the 300V spectra. In all 1200R spectra, the Na\,{\sc i}\,{\sc d} doublet is resolved separately as narrow emission lines at 5889 and 5895 \ang (each with FWHM$\approx$220 \kms). These are emitted at the rest wavelength in the host galaxy. There is no sign of He\,{\sc i} $\lambda5876$ in emission. Multiple absorption components are visible at $\sim$5640, $\sim$5755 and $\sim$5860 \ang at $+$35 days. We note that the feature around 5600-5900 \ang shows a similar structure to the He\,{\sc i} $10830$ \ang feature observed by \citet{Fra2013}. This similarity leads us to propose that the contribution of Na\,{\sc i}\,{\sc d} to the broad line feature is minor. The He\,{\sc i} absorption is strongest at $\sim$5640 \ang with velocity $\sim-$11800 \kms, the other two components are observed at a velocity of $-$6200 and $-$800 \kms. The absorption components become stronger at $+$42 days and a more complex velocity structure emerges, with further narrow absorption components appearing. The low-velocity ($<$ 5000 \kms) absorption components increase in strength by $+$64 days, so that the feature appears to be composed of two separate absorption components at $\sim$$-$4000 \kms and $\sim  -$12000 \kms. Note that the higher velocity component may be Sc {\sc ii} $\lambda 5658$ in absorption, as has been identified in SN 1999em by \citet{Elm2003}, Sc {\sc ii} $\lambda 5526$ is also seen in emission at $+$64 and $+$73 days. At $+$64 and $+$73 days the narrow emission of Na\,{\sc i}\,{\sc d} is replaced by a broader emission resulting from the strengthening of the He\,{\sc i} $\lambda5876$ emission line. This is supported by the increased strength of the  He\,{\sc i} $\lambda\lambda$ 5015, 6678 and 7065 \ang lines at those epochs.
\\
\indent The Ca\,{\sc ii} infrared (IR) triplet appears as three weak narrow emission lines with a potential underlying broad emission at $+$35 and $+$42 days, by the third epoch the emission lines become more pronounced with associated blueward broad absorption troughs ($\sim -4000$ \kms).
  

\section{Analysis of the polarimetry}
\label{sec:analysis}
The polarisation spectra of SN 2009ip from 7 November 2012 ($+$35 days) to the 25 December 2012 ($+$83 days) as observed with the 300V and 1200R grisms are presented in Figures \ref{300vpolflux} and \ref{1200rpolflux} below, respectively. The flux spectra are also shown in order to aid the identification of polarised signals associated with particular line features.

\begin{figure*}
\includegraphics[width=18cm]{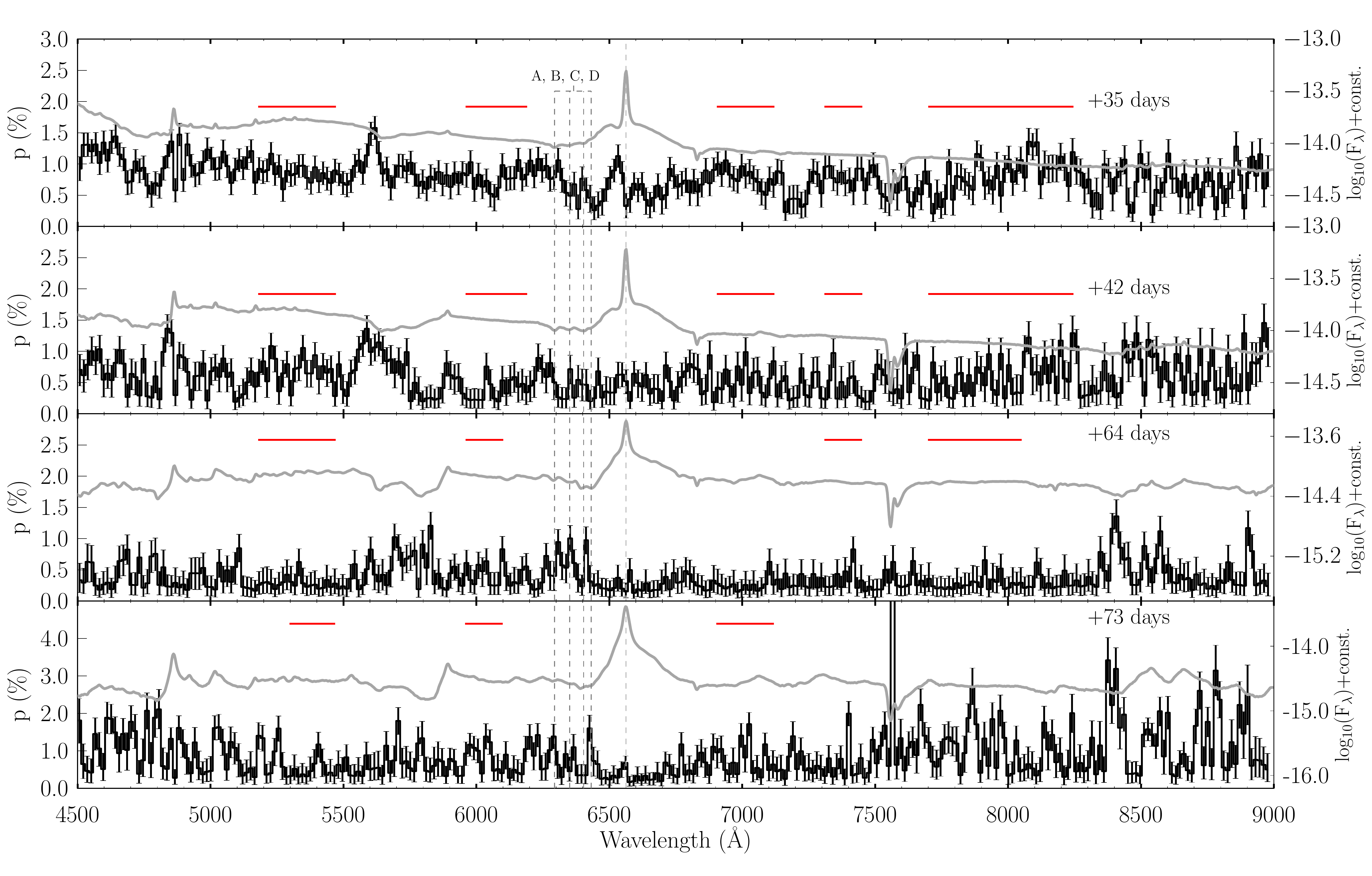}
\caption{Spectropolarimetric data taken with the 300V grism starting on 7 November 2012 and ending on 15 December 2012. The degree of polarisation as a function of wavelength is presented with the data binned to 15 \ang and corrected for the ISP. The flux spectra are in grey with units $\mathrm{10^{-14}\, erg s^{-1} cm^{-2} \AA ^{-1}}$ and are corrected for the recessional velocity of the host galaxy, $cz=1782\pm5$ \kms \citep{Mey2004}. The wavelength regions that were used to calculate the continuum polarisation are shown by the red horizontal bars. Note the different scaling in each panel in flux and the degree of polarisation.}
\label{300vpolflux}
\end{figure*}

\begin{figure*}
\includegraphics[width=18cm]{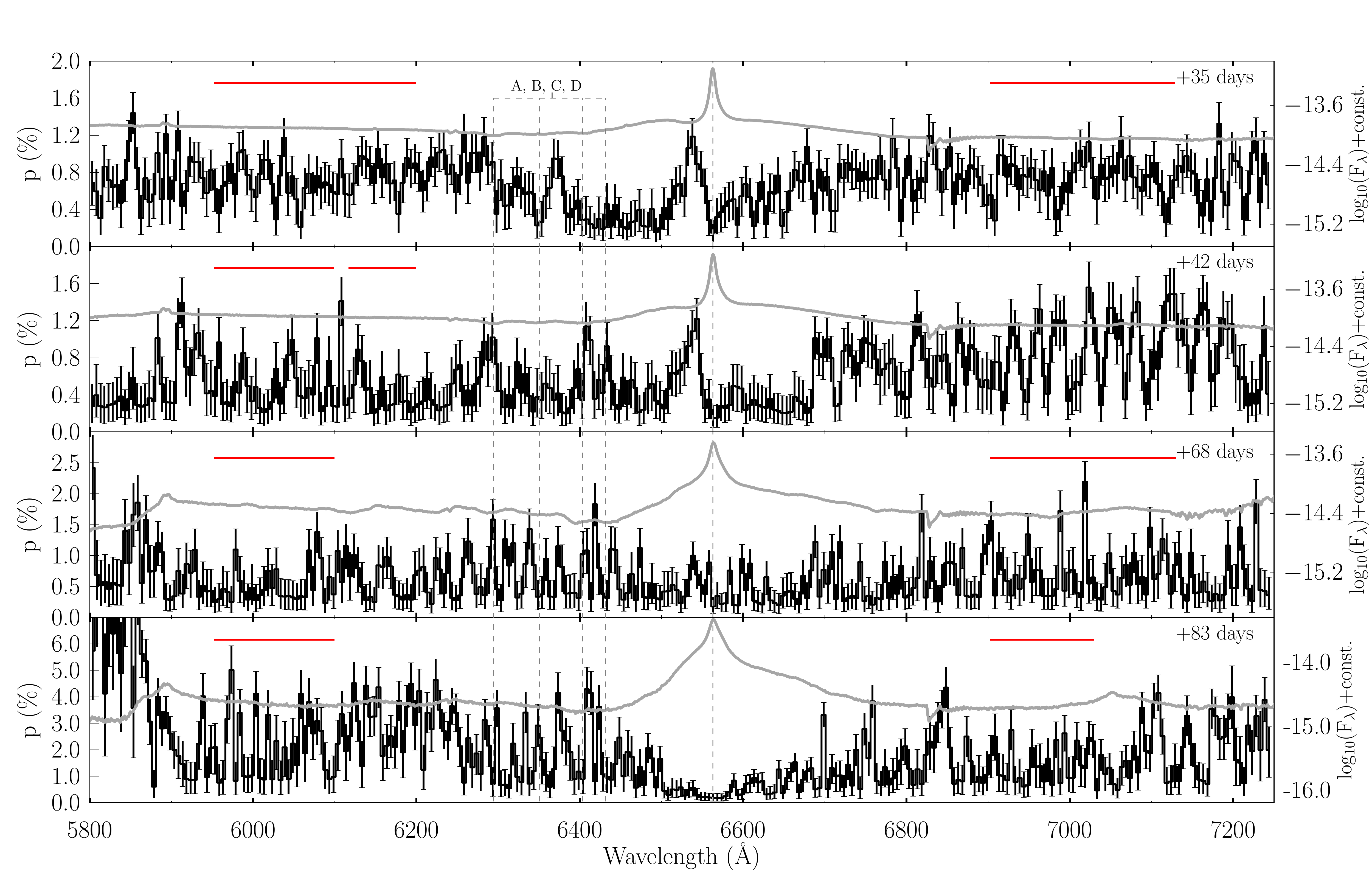}
\caption{Same as Figure \ref{300vpolflux}, but for spectropolarimetric data taken with the 1200R grism starting on 7 November 2012 and ending on 25 December 2012.}
\label{1200rpolflux}
\end{figure*}

\begin{figure*}
\includegraphics[scale=0.43]{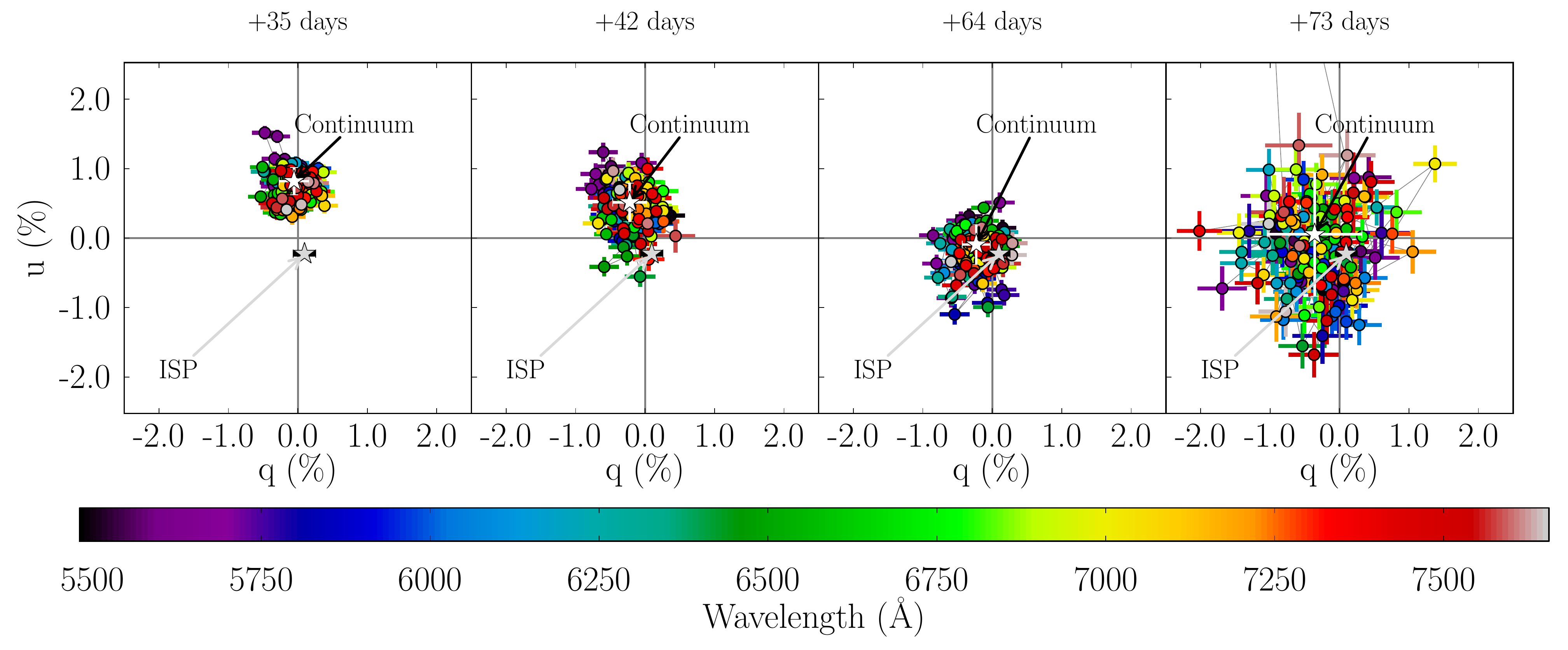}
\includegraphics[scale=0.43]{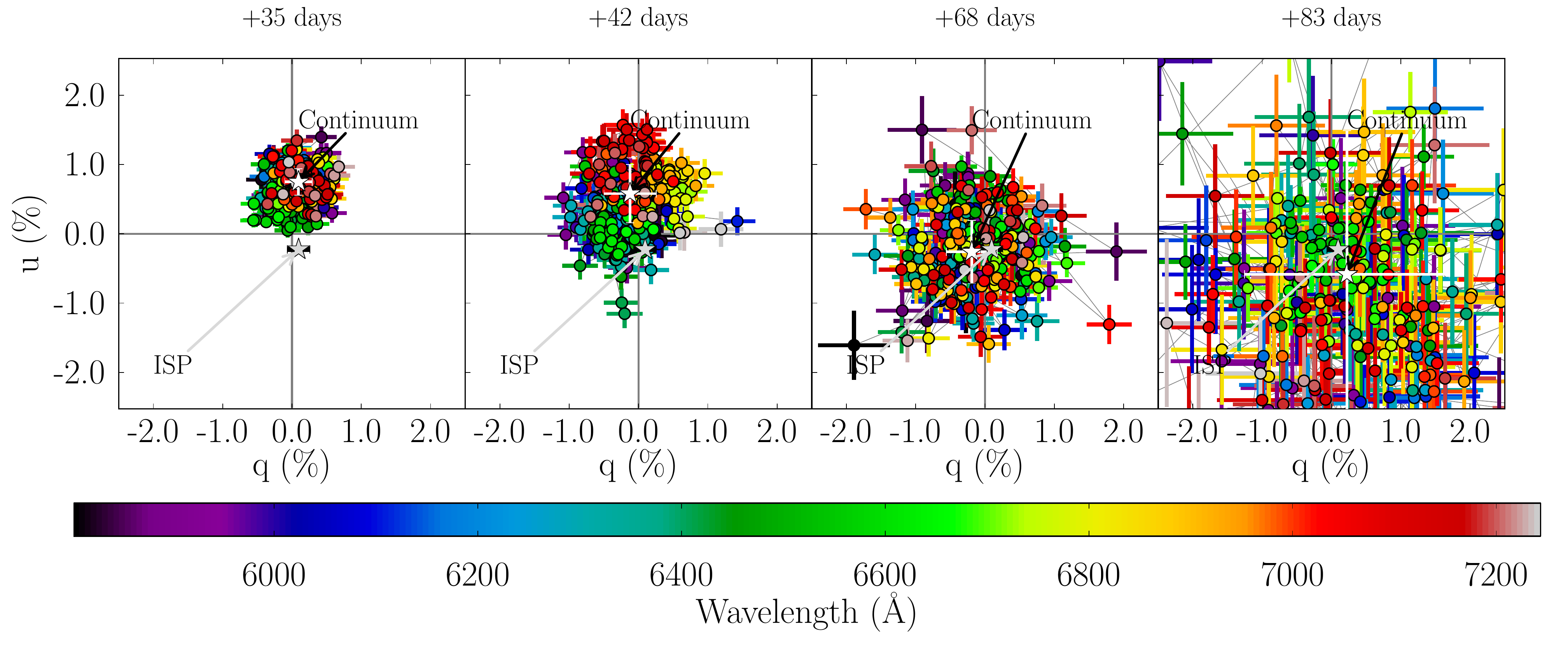}
\caption{The evolution of SN 2009ip on the $q$-$u$ plane. The top panel contains the data from the 1200R observations (binned to 5 \ang or $\sim 230$\kms)  and the bottom panel the observations taken with the 300V grism (binned to 15 \ang or $\sim 680$\kms). The position of the continuum is marked by the white star with the 1$\sigma$ error bars. The data are corrected for the interstellar polarisation shown as a grey star with black error bars. }
\label{qu}
\end{figure*}

\subsection{Interstellar polarisation} 
\label{subsec:ISP}

Photons scattering from gas and dust grains in the interstellar and circumstellar material introduce a source of polarisation in the spectrum that is not intrinsic to the transient. The removal of the interstellar polarisation (ISP) is therefore vital to the analysis so that we may understand the geometry of the event. 

The foreground Galactic reddening for the Milky Way Galaxy (MW) along the line of sight to SN 2009ip is $E(B-V) = 0.019$ mags \citep{Sch1998} \footnote{http://ned.ipac.caltech.edu}. The interstellar polarisation expected for the MW dust component is $<0.17\%$ \citep{Ser1975}. The low polarisation is corroborated by that of a standard star, HD 212146,  40' from the position of SN 2009ip, which has $p=0.08\pm0.04\%$ \citep{hei2000}; however, the distance to this star is only $\sim40\,\mathrm{pc}$ \citep{van2007}, such that it only provides a lower limit on the Galactic ISP arising along the line of sight.

The component of the ISP from the host galaxy is also expected to be small, as NGC 7259 is viewed face-on and SN 2009ip is positioned at the extreme edge of the optical disk at 4.3kpc \citep{Fra2013}. The data (Figures \ref{300vpolflux} and \ref{1200rpolflux}) show no strong wavelength dependence in the continuum polarisation, indicating that the ISP is small. A straight line fit with a least squares method to the Stokes $q$ and $u$ spectra at $+$35 days with a 1$^{\mathrm{st}}$ order polynomial yielded a weak dependence on the wavelength of $du/d\lambda=+1.5\pm1.9\times10^{-7}$ /\ang and $dq/d\lambda=+4.9\pm2.0\times10^{-7}$ /\ang.

A value for the ISP was determined assuming that the emission of \halpha\,\,was intrinsically unpolarised, such that the line polarisation ($q_{\rm line}$ and $u_{\rm line}$) at these wavelengths should reflect the ISP. Indeed, there is evidence of depolarisation associated with the narrow emission line at all epochs and with the broad wings at $+$35 and $+$42 days. Two estimates of the line polarisation were determined under different assumptions: 
\begin{enumerate}
  \item ISP$_{A}$ assumed that the broad emission of \halpha\,\,under the narrow emission component was unpolarised;
  \item ISP$_{B}$ assumed that only the narrow emission component was unpolarised.
\end{enumerate}

The Stokes line polarisation parameters, $q_{\rm line}$ and $u_{\rm line}$ are defined as follows; 
\begin{equation}
q_{\rm line}=\frac{q_{\rm tot}F_{\rm tot}-q_{\rm base}F_{\rm base}}{F_{\rm line}}\\
\end{equation}
\begin{equation}
u_{\rm line}=\frac{u_{\rm tot}F_{\rm tot}-u_{\rm base}F_{\rm base}}{F_{\rm line}}\\
\end{equation}

These equations assume that, at any given wavelength, the observed total polarisation is the sum of the line polarisation (if any) and some base polarisation that needs to be removed from the observed polarisation to determine the line polarisation. The flux of each component is denoted by $F$.  

For ISP$_{A}$, a flat continuum polarisation was taken as the base polarisation. The continuum polarisation, $q_{\rm cont}$ and $u_{\rm cont}$, was determined as the weighted average of Stokes $q$ and $u$ in the wavelength regions indicated in Figure \ref{1200rpolflux}. The continuum flux, $F_{\rm cont}$ was measured from the bottom of the broad \halpha\,\,wings. In this case $q_{\rm base}=q_{\rm cont}$, $u_{\rm base}=u_{\rm cont}$ and $F_{\rm base}=F_{\rm cont}$ in the above equations.

For ISP$_{B}$, the depolarisation observed in Stokes $q$ and $u$ associated with the broad wings of \halpha\,\,at $+$35 and $+$42 days, were approximated with a Gaussian profile. The line intensity was measured from the base of the narrow \halpha\,\,P Cygni profile. To determine the line polarisation in this case, the continuum polarisation and the broad \halpha\,\,component were accounted for such that
\begin{equation}
q_{\rm base}=q_{\rm cont}+q_{\rm broad}
\end{equation}
and similarly for $u_{\rm base}$ and:
\begin{equation}
F_{\rm base}=F_{\rm cont}+F_{\rm broad}
\end{equation}
  The $q_{\rm line}$ and $u_{\rm line}$ of a 5 \ang bin centred on the narrow \halpha\,\,emission in the 1200R spectra were taken as the estimate of the ISP in each epoch. The higher spectral resolution 1200R data were chosen to avoid contamination from potential polarisation in the broad wings of \halpha. The value adopted as the ISP was then taken as the inverse error weighted average of those from the first 3 of the 4 epochs of 1200R datasets presented. The signal-to-noise ratio of the polarisation spectrum at $+$83 days was considered too low to provide an accurate determination of the ISP. Averaging the measurements across multiple epochs makes the best use of the data, as the ISP should not vary with time.
    
ISP$_{A}$ was found to be $q_{\rm ISP}$=0.10$\pm$0.16$\%$ and $u_{\rm ISP}$=$-$0.224$\pm$0.019$\%$, corresponding to $p_{\rm ISP}$=0.18$\pm$0.13$\%$ and $\theta_{\rm ISP}$=146$\pm$22$^{\circ}$. This is consistent with the upper limit derived from the reddening above. Under the assumption that the broad emission was polarised, ISP$_{B}$ was determined to be $q_{\rm ISP}$=0.30$\pm$0.09$\%$ and u$_{\rm ISP}$$=-$0.17$\pm$0.06$\%$, corresponding to $p_{\rm ISP}$=0.33$\pm$0.09$\%$ and $\theta_{\rm ISP}$=164$\pm$8$^{\circ}$. The two estimates for the ISP agree to within their errors. The uncertainties on $q_{\rm ISP}$ and $u_{\rm ISP}$ are the standard deviation of the individual measurements from the three epochs around the final inverse error weighted average.

 As ISP$_{A}$ relied on fewer assumptions, this was the one adopted as the ISP. The ISP lies close to the origin on the $q$-$u$ plane, in the positive $q$, negative $u$ quadrant, as marked in Figure \ref{qu}. As there is no discernible wavelength dependence in the observed continuum polarisation at all epochs, we apply a single correction for the ISP to the entire wavelength range. ISP$_{A}$ was subtracted from Stokes $q$ and $u$ and the uncertainties propagated by addition in quadrature, $p$ and $\theta$ were then recalculated. 

\subsection{The intrinsic continuum polarisation}
\label{subsubsec:cont}
\begin{figure}
\centering
\includegraphics[scale=0.30]{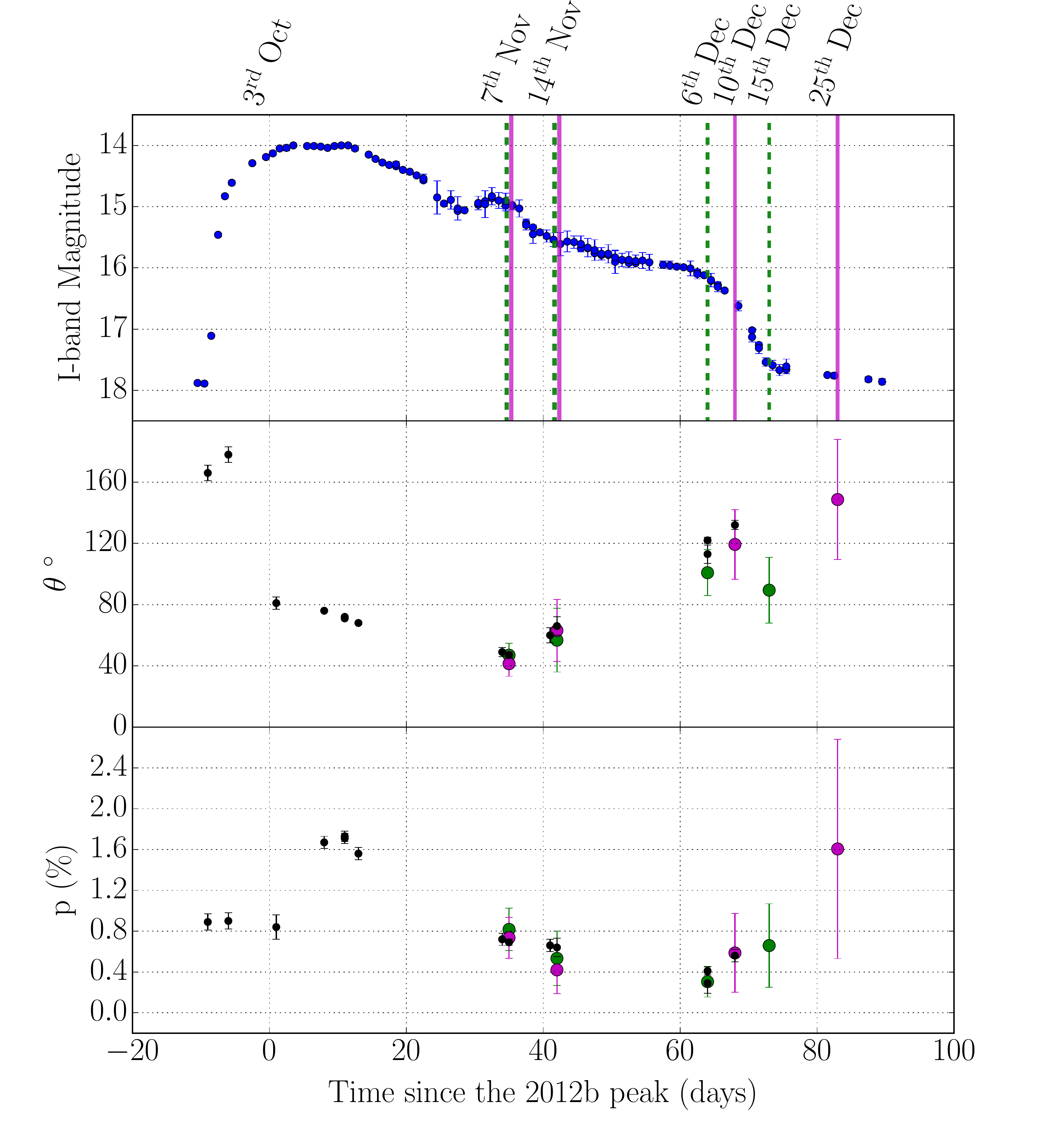}
\caption{The evolution of the continuum polarisation over the course of the 2012b outburst. The top panel shows the $I$-band light curve using data presented by \citet{Gra2014} with the dates of the polarimetry observations overlaid with the dashed green lines for the 300V data and solid magenta lines for the 1200R data. The polarisation angle as a function of time is shown in the middle panel, followed by the polarisation degree in the bottom panel. The smaller black markers represent the $V$-band polarisation measurements as presented by \citet{Mauer2014}. }
\label{continuumvtime}
\end{figure}

Upon removal of ISP$_{A}$, the continuum polarisation was determined by taking the inverse error weighted average of Stokes $q$ and $u$ in the wavelength regions specified in Figures \ref{300vpolflux} and \ref{1200rpolflux}. Wavelength regions where the polarisation was approximately flat and where there were no obvious line features in the corresponding flux spectrum were chosen to determine the continuum polarisation. The uncertainty of the Stokes $q$ and $u$ of the continuum corresponds to the standard deviation in the chosen regions. The degree and angle of polarisation, $p$ and $\theta$, for the continuum were then calculated, with $p$ de-biased as described in Section \ref{sec:obs}. These are presented in Table \ref{continuumtable}. 

The continuum is significantly polarised in the first epoch at $p\sim$0.7\% and $\theta \sim$ 45$^{\circ}$. By the second epoch of $+$42 days, the degree of polarisation has slightly decreased and is accompanied by a small rotation in the polarisation angle. There is no significant change in the degree of polarisation between $+$42 days and $+$64 days, there is, however, a substantial change in the angle which undergoes a rotation of $\sim$60$^{\circ}$. Following this, both the degree and angle of polarisation remain approximately constant at the 1$\sigma$ level throughout the rest of the observations. The uncertainty in $p$ and $\theta$ increases in the later epochs due to the decreasing signal as the transient fades. 

The rotation of the polarisation angle by the third epoch is also clear from the evolution on the $q$-$u$ plane (Figure \ref{qu}). The large rotation occurs at the same time as a drop in the bolometric light curve as reported by \citet{Mar2014}. The change in the orientation of the polarisation is possibly related to this event. The $I$-band light curve produced by \citet{Gra2014} is shown in Figure \ref{continuumvtime} and also shows the drop in flux at $\sim+$70 days. The $I$-band light curve was chosen to display as the drop at $\sim+$70 days is most obvious. The robustness of the observed rotation was tested by changing the determined ISP$_{A}$ to its upper and lower 3$\sigma$ boundaries. The measured rotation was found to be partially dependent on the choice of ISP, with the rotation decreasing from $\sim$60$^{\circ}$ to $\sim$20-40$^{\circ}$. We consider it robust that the continuum polarisation rotates, however the error introduced by the choice of ISP implies that the degree to which it rotates should be approached with caution.

\begin{table}
\centering
\caption{The ISP-corrected continuum polarisation of SN 2009ip.  }
 \begin{tabular}[alignment]{c c c c c c c}
 Phase (days)$^{\dag}$ & p (\%) & $\theta$ ($^{\circ}$) \\
 \hline
  & 300V &\\
  \hline
 $+$35 & 0.76$\pm$0.21 & 47$\pm$8 \\
 $+$42 & 0.37$\pm$0.30 & 57$\pm$21 \\
 $+$64 & 0.29$\pm$0.24 & 100$\pm$15\\
 $+$73 & 0.66$\pm$0.55 & 89$\pm$21 \\
   \hline
   & 1200R & \\
   \hline
 $+$35 & 0.69$\pm$0.20 & 41$\pm$8 \\
 $+$42 & 0.33$\pm$0.26 & 63$\pm$20 \\
 $+$68 & 0.48$\pm$0.39 & 119$\pm$22\\
 $+$83 & 0.78$\pm$1.07 & 149$\pm$39 \\

  \end{tabular}
  \vspace{2mm}
    \begin{flushleft}
  \dag Relative to UV-band maximum of the 2012b re-brightening on the 3$^{\rm rd}$ October 2012\\
  \end{flushleft}
  \label{continuumtable}
\end{table}

\subsection{Intrinsic line polarisation}
\label{sec:linepol}
The flux spectra of SN 2009ip exhibit complex line features, with narrow and broad P Cygni profiles and absorption notches embedded at both low and high velocity. Similarly, the polarisation spectra also show a complex structure, with some of the features common to multiple species. In the following subsections we separate these features according to the breadth of the profile (narrow and broad features) and the velocity at which they appear, with respect to the rest wavelength of the line with which they are associated. The velocities, degree and angle of polarisation of all the features summarised in the following subsections can be found in Table \ref{table:linepol} in the Appendix.

\subsubsection{Broad polarised features: \halpha}
\label{subsubsec:broadhalpha}

\begin{figure}
\includegraphics[scale=0.33]{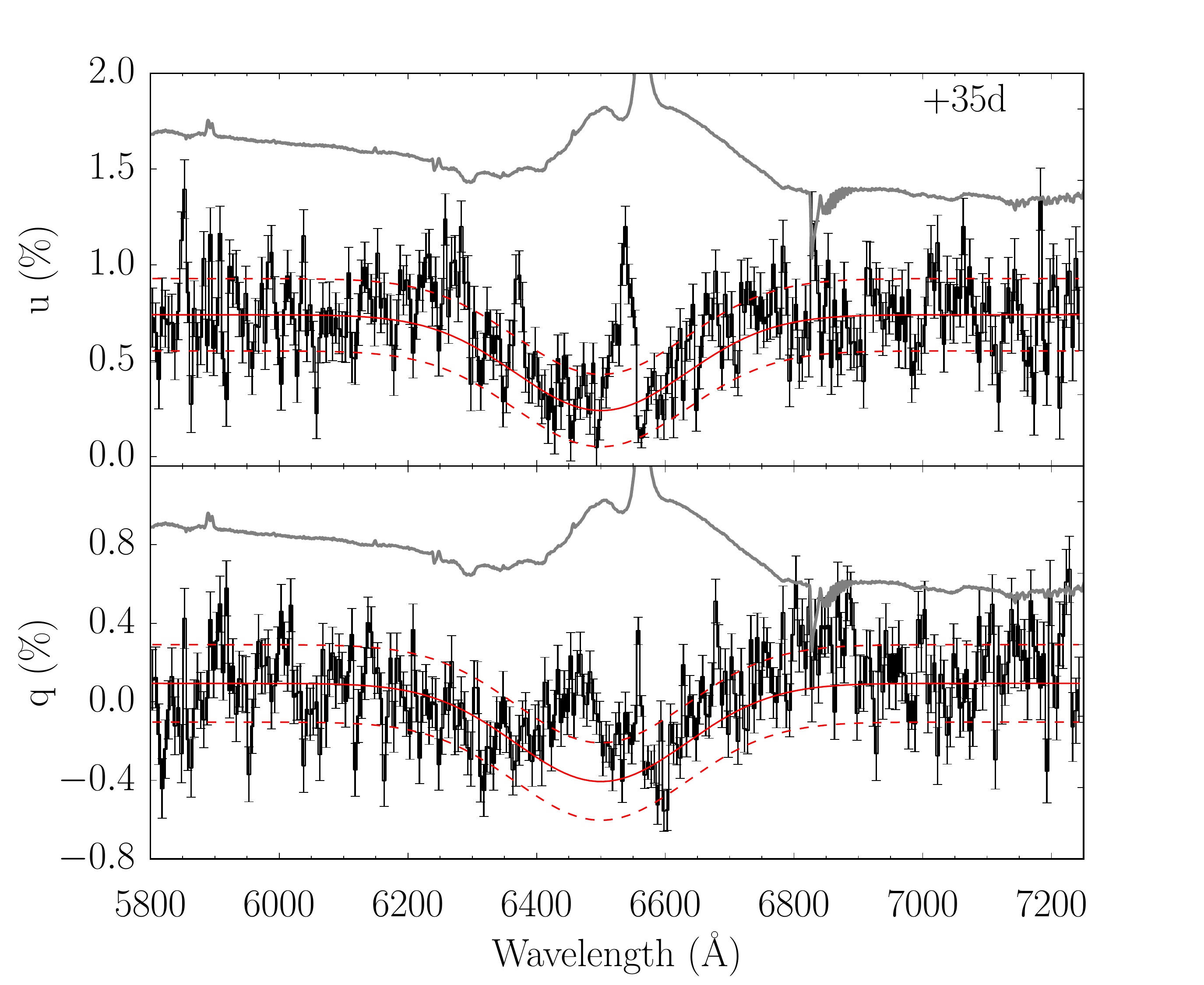}
\caption{The fit to the broad \halpha\,\,depolarisation in $q$ and $u$ at $+$35 days as observed with the 1200R grism. The top and bottom panels show $q$ and $u$ respectively along with their Gaussian fit to the depolarisation (the $\pm$1$\sigma$ fits are shown by the dashed red lines), used to isolate the polarisation of the low-velocity narrow \halpha\,\,line.}
\label{1107curvedcont}
\end{figure}

In the flux spectrum, \halpha\,\,is observed as a combination of a narrow P Cygni profile superimposed on a broad P Cygni profile. There is obvious depolarisation associated with the broad emission ($\sim 10^{4}$ \kms) at $+$35 and $+$42 days, as can be seen in Figures \ref{300vpolflux} and \ref{1200rpolflux}. No other species in the spectra shows the same broad emission or depolarisation at these epochs, which could be a consequence of the strength of the line.

For data taken at $+$35 and $+$68 days, the observed polarisation, $p_{\rm obs}$, of the broad \halpha\,\,emission in 1200R was modelled with the intrinsic line polarisation, $p_{\rm line}$, as a variable. The model took the form of:
\begin{equation}
p_{\rm obs}=\frac{p_{\rm line}F_{\rm line}+p_{\rm cont}F_{\rm cont}}{F_{\rm obs}}
\end{equation}
where the intensity of the continuum, $F_{\rm cont}$ was fit with a second order polynomial. Two wavelength regions were chosen in order to determine whether the intrinsic polarisation, $p_{\rm line}$, of the broad emission was fixed or variable as a function of wavelength.  These were 6600 - 6700 \ang (corresponding to $\sim$1700 - 6300 \kms) for the core and 6700 - 6800 \ang ($\sim$6300 - 10000 \kms) for the outer red wing of the line profile.

At $+$35 days, the observed polarisation appears to be best fit with $p_{\rm line} \leq 0.3 $\% in the core. The $\chi$$^{2}_{\nu}$ was minimised (to a value of $1.68$) for $p_{\rm line}=0.2 \pm 0.1$\%. The improvement in $\chi$$^{2}_{\nu}$ for $p_{\rm line}$ = 0.2\% over $p_{\rm line}$=0.0 \% is only fractional at 0.12. The residuals of the fit of $p_{\rm obs}$ to the observed $p$ are also evenly scattered for $p_{\rm line} \leq 0.3 $\%. The line polarisation in the outer wings, however, is not as well constrained. The $\chi^{2}_{\nu}$ is minimised to a value of 1.38 by $p_{\rm line}$=0.5$\pm$0.1\%. This corresponds to only a 0.043 improvement in the reduced chi squared over either $p_{\rm line}$=0.0\% or $p_{\rm line}$=0.9\%. The data in this region appear to be equally well represented by either of these values. In this case it is difficult to discern if the outer wings are more polarised than the core of the broad emission.

At $+$68 days, $\chi^{2}_{\nu}$ is minimised for $p_{\rm line}=0.3\pm0.1\%$ for both wavelength regions. This represents a 0.25 improvement in $\chi^{2}_{\nu}$ in the core over $p_{\rm line}=0.0\%$ but only a 0.06 improvement for the outer wings. The data appear to be well represented by $p_{\rm line}\leq0.5\%$, with the increased scatter in the polarisation meaning that it can be constrained no further. 

The polarisation blueward of the narrow emission is not well represented by the fits applied to the redward broad emission. This is also clear in Figure \ref{1107curvedcont}, where the Stokes parameters across \halpha\,\,at $+$35 days are presented. The data show narrow (mostly in $u$) departures from the fit at $\sim$6370 \ang ($v\sim-$8820 \kms) and $\sim$6540 \ang ($v\sim-$1000 \kms), these will be analysed in the following subsections. A broader ($\sim$ 5500 \kms) deviation from the fit is observed mostly in Stokes $q\sim0.2-0.6\%$ peaking at $\sim$6460 \ang. This is indicative of polarisation intrinsic to the broad P Cygni absorption.

The polarisation angle of the broad component blueward of \halpha\,\,was found using the 1200R spectra at $+$35, $+$42 and $+$68 days. An inverse error weighted average of $q_{\rm line}$ and $u_{\rm line}$ was calculated in regions avoiding the narrow polarisation features associated with the absorption notches A, B, C and D. The feature between 6300-6500 \ang ($\sim-$12000 - $-$3000 \kms) was found to be polarised to the $p_{\rm line}\sim$0.3-0.65\% level from $+$35 to $+$68 days. The polarisation angle was consistent at $\theta_{\rm line}\sim$125$^{\circ}$ in the first two epochs before rotating to 85$\pm$42$^{\circ}$, the large uncertainty on the latter value means that it is also consistent with the previous values to within 1$\sigma$. 

\subsubsection{Low-velocity narrow polarisation}
\label{subsubsec:narrowhalpha}

\begin{figure}
\includegraphics[scale=0.31]{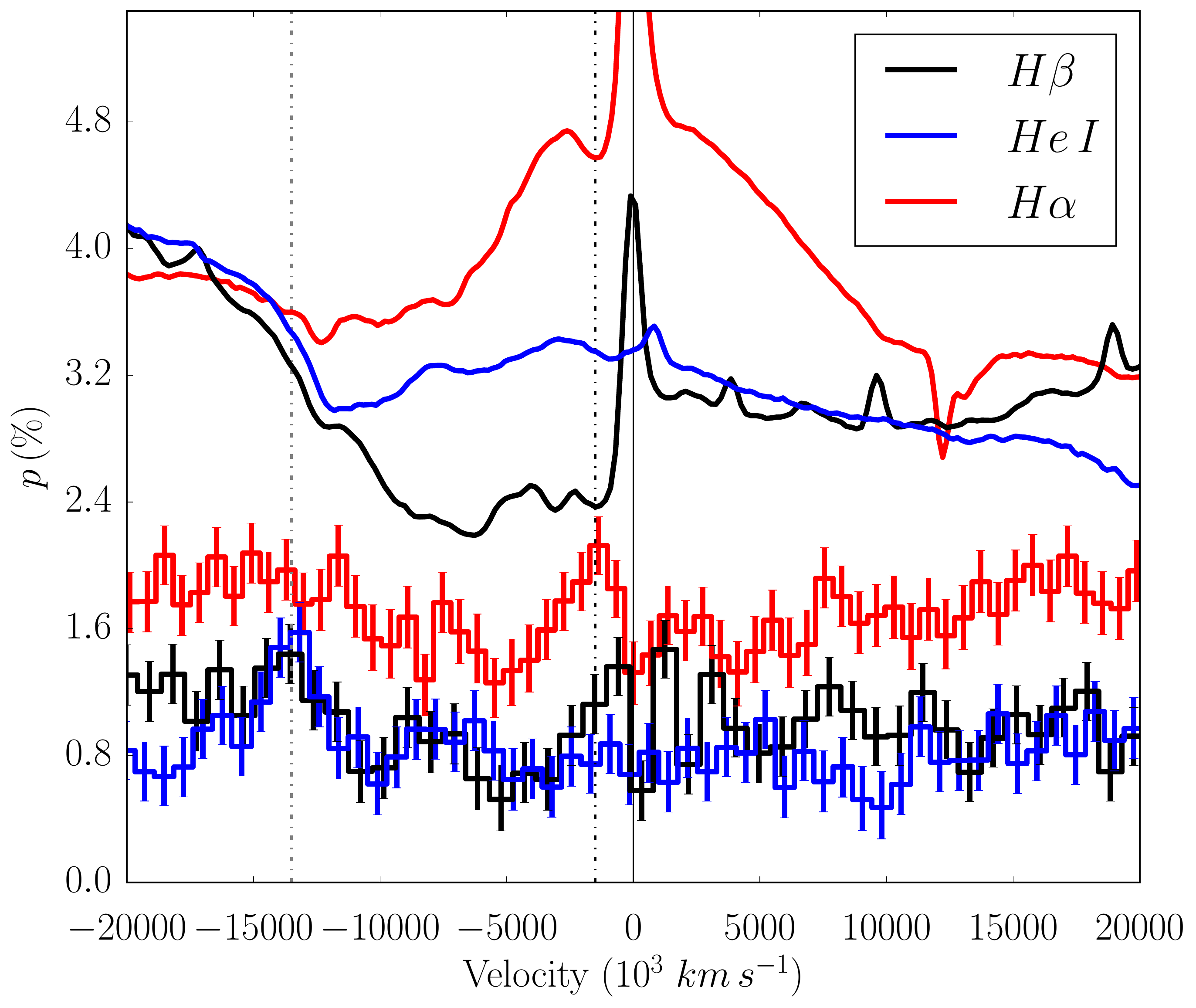}
\caption{Line profiles and 300V polarisation spectra for \halpha, \hbeta\,\,and He\,{\sc i} $\lambda 5876$ in velocity space at $+$35 days.The velocity of the absorption component (at absorption minimum) of the narrow P Cygni profile of \halpha\,\,is indicated by the black dashed-dotted vertical line. The grey dash-dotted line indicates the position of significant polarisation observed for \hbeta\,\,and He\,{\sc i} $\lambda 5876$ at $\sim-$14000 \kms. Peaks in the polarisation can also be seen at low velocities ($\sim-$1000 \kms) for \halpha\,\,and \hbeta. Note that the polarisation spectrum across \halpha\,\,(in red) has been offset by 1\% for clarity.}
\label{halphahbeta300V}
\end{figure}

\begin{figure}
\includegraphics[scale=0.31]{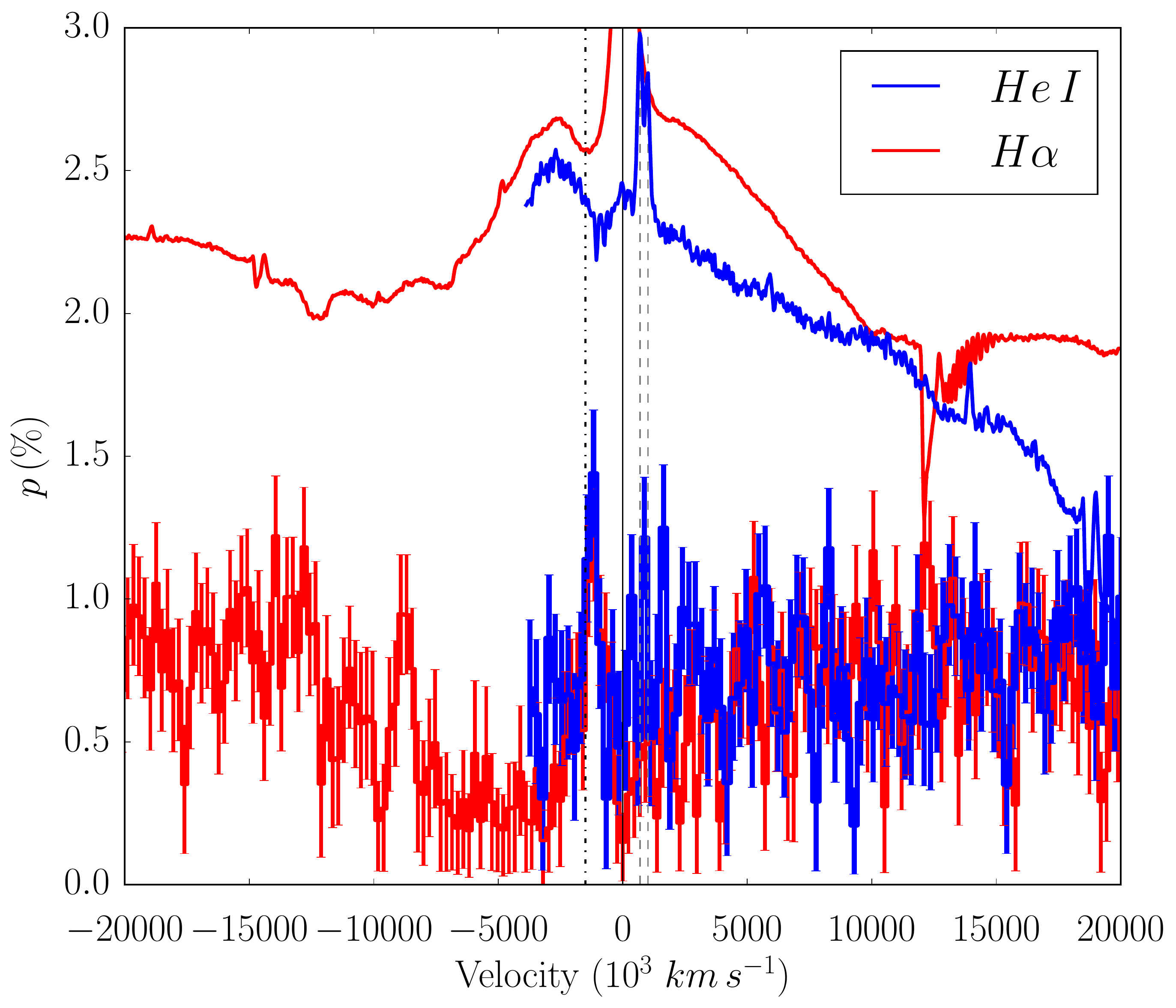}
\caption{Same as Figure \ref{halphahbeta300V} but for \halpha\,\,and He\,{\sc i} $\lambda 5876$ as observed with the 1200R grism at $+$35 days. The position of the Na\,{\sc i}\,{\sc d} emission lines are marked by the grey dashed lines. Again the velocity at absorption minimum for the narrow P Cygni profile of \halpha\,\,is indicated by the black dash-dot vertical line. A peak in the polarisation is observed at $\sim-$1000 \kms in both hydrogen and helium, implying a shared structure at low velocities for these line-forming regions. The He\,{\sc i} profile is truncated as it fell at the blue edge of the wavelength range of the 1200R observations.}
\label{halphahbeta1200R}
\end{figure}

\begin{figure*}
\centering
\includegraphics[scale=0.4]{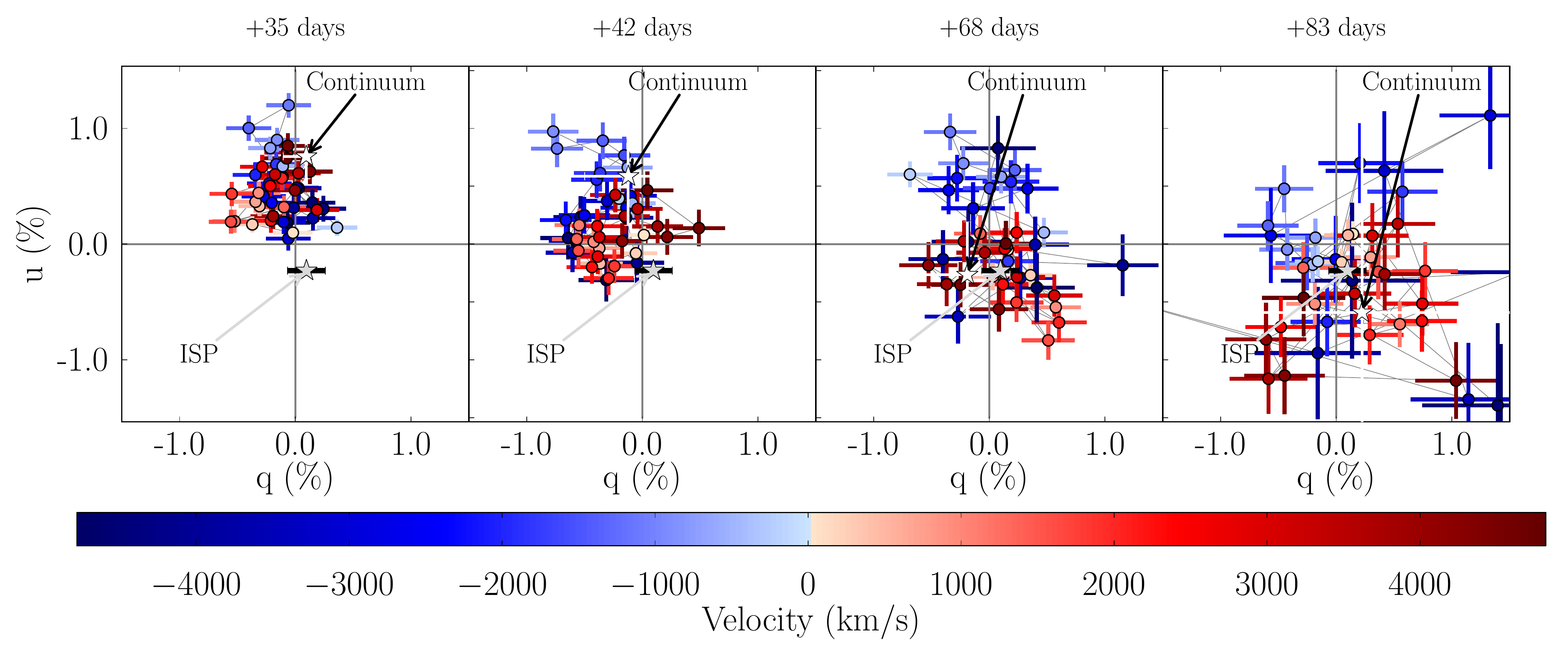}
\caption{The evolution of \halpha\,\,on the $q-u$ plane as observed with the 1200R grism in place using 5 \ang ($\sim 230$\kms) binning. The position of the continuum is marked by the white star with the 1$\sigma$ error bars. The position of the ISP is indicated by the grey star. Stokes $q$ and $u$ have been corrected for the ISP in these plots.}
\label{halphaqu}
\end{figure*}

It can be seen from Figure \ref{halphahbeta300V} that \halpha, \hbeta\,\,and He\,{\sc i} $\lambda 5876$ have complicated line profiles with multiple absorption components at both high and low velocity. In particular, they each show a narrow absorption component at $\sim-$1500 \kms. In the polarisation spectra of $+$35 days, presented in Figure \ref{halphahbeta300V}, a large peak is observed associated with this component for \halpha\,\,and \hbeta. This feature is also observed in the 1200R data for \halpha\,\,and He\,{\sc i} $\lambda 5876$, as seen in Figure \ref{halphahbeta1200R}. For \halpha, the feature is evident on the $q$-$u$ plane in Figure \ref{halphaqu} at $+$42 days. It can be seen as a small number of data points in the upper left at low velocities ($\sim$$-$1000 \kms) that protrude away from where the continuum is indicated. The \halpha\,\,peak is visible until $+$68 days in the 1200R data but is not polarised by more than 3$\sigma$ after $+$35 days in the 300V data. The difference in the strength of the feature between the two datasets is largest at $+$42 days. Smoothing the 1200R data to the resolution of the 300V grism and using the same bin size resolved this discrepancy. This highlights the need for high-resolution spectropolarimetric observations for narrow line SNe. The degree of polarisation decreases over time as the strength of the absorption in the flux spectrum decreases.

As the spectra contain a number of narrow features that are marginally significant we have employed a two-tailed $z$-test across our data \citep{Spr2011}. We only analyse those features where the two-tailed $z$-test gives a probability of more than $\sim3\sigma$ (having a probability less than 0.27\%) that the feature is not consistent with the continuum. The depolarisation associated with the broad \halpha\,\,component was accounted for when calculating the z-test statistic for the low-velocity narrow \halpha\,\,signal and the high-velocity narrow ``notches". 
 
In order to isolate the line polarisation of the low-velocity \halpha\,\,feature, the depolarisation of the broad \halpha\,\,emission component was approximated with a Gaussian profile. This was then vector subtracted from the observed data. The fit to the 1200R Stokes $q$ and $u$ at $+$35 days is shown in Figure \ref{1107curvedcont}. This was fit for both the 1200R and 300V spectra at $+$35 and $+$42 days. For the later epochs, when there is no discernible depolarisation, only the flat continuum polarisation, as determined in Section \ref{subsubsec:cont}, was subtracted.

In the 1200R spectrum of $+$35 days, the degree and angle of polarisation of the low-velocity feature for \halpha\,\,and He\,{\sc i} $\lambda 5876$ are consistent at $p_{\rm line}\sim$1\% and corresponding angle of $\theta_{\rm line}\sim$33$^{\circ}$. The signals display a striking similarity when compared in velocity, as seen in Figure \ref{halphahbeta1200R}. For \hbeta\,\,the degree of polarisation, as revealed in the 300V spectrum in Figure \ref{halphahbeta300V}, at this epoch is observed to be lower at $p_{\rm line}$=0.4$\pm$0.3\% and $\theta_{\rm line}$=44$\pm$24$^{\circ}$. The polarisation of the same low-velocity \halpha\,\,feature is also weaker in the 300V data at $p_{\rm line}$=0.30$\pm$0.23\% and the angle slightly rotated and with a larger uncertainty, $\theta_{\rm line}$=50$\pm$15$^{\circ}$. Similarly, the He\,{\sc i} $\lambda 5876$ polarisation is not observed in the 300V spectrum taken on the same night (this is evident when Figures \ref{halphahbeta300V} and \ref{halphahbeta1200R} are compared), demonstrating the loss of information when the spectra are binned to 15 \ang.

At $+$35 days, the low-velocity polarised peaks of \halpha\,\,and \hbeta\,\,are redshifted with respect to the absorption line minimum observed in the flux spectrum (see Figure \ref{halphahbeta300V}). In the 1200R spectrum, this is by 350 \kms for \halpha, where for He\,{\sc i} $\lambda 5876$ the polarised peak is blueshifted from the absorption minimum. It is possible that blending with Na\,{\sc i}\,{\sc d} has smeared out the total flux line profile. Given the similarity in the polarised signals in velocity space, it would appear that Na\,{\sc i}\,{\sc d} does not significantly contribute to the polarised line profile.

The \halpha\,\,signal remains polarised through to $+$68 days where it is detected at 2.8$\sigma$ (according to the z-test) and exhibits no significant change in the polarisation angle. In contrast, the low-velocity He\,{\sc i} $\lambda 5876$ feature is not observed after $+$35 days. The \hbeta\,\,signal, conversely, increases in degree of polarisation to $p_{\rm line}$=1.1$\pm$0.3\% and undergoes a rotation to 98$\pm$9$^{\circ}$ before dissipating by $+$64 days. The similarity in the polarisation of \halpha, \hbeta\,\,and He\,{\sc i} $\lambda 5876$ at low velocities indicates that hydrogen and helium share an asymmetric slowly moving ($-$1100 \kms) line-forming region.

\subsubsection{Polarisation at intermediate and high velocities}
\label{subsubsec:highvnarrow}

Multiple species show peaks in the polarisation at intermediate and high velocity associated with the observed absorption components (see Figure \ref{halphahbeta300V}). Similarly to the low-velocity feature discussed in the previous subsection, some polarised features have corresponding signals associated with another line or species, indicating that they may arise in similar line-forming regions. This, however, is not the case for all detected line polarisation. Further information of the polarisation of features summarised here can be found in Table \ref{table:linepol} in the Appendix.

At $+$35 days, He\,{\sc i} $\lambda 5876$ and \hbeta\,\,display peaks at $\sim-$14000 \kms with $p_{\rm line}\sim$0.5-0.7\%  and consistent polarisation angle of $\theta_{\rm line}\sim$55$^{\circ}$. Figure \ref{halphahbeta300V} shows that the profiles are also similarly shaped in velocity space. Both polarisation peaks are also blueshifted from the nearest absorption minimum by $\sim$2000 \kms. Similarly, at $+$64 days, \halpha\,\,and He\,{\sc i} $\lambda 5876$ share a peak of strength $p_{\rm line}$=0.7$\pm$0.3\% and angle of $\theta_{\rm line}\sim$120$^{\circ}$  at $-$9800 \kms. At $+$73 days, \hbeta\,\,and He\,{\sc i} $\lambda 5876$ again show comparable polarisation at $-$6350 \kms at a similar position in the sky. In flux, the velocity structure of \halpha, \hbeta\,\,and He\,{\sc i} $\lambda 5876$ are similar as shown in Figure \ref{halphahbeta300V}; the structure in polarisation corroborates this. The data imply that at high velocities as well as low, hydrogen and helium share the same asymmetrical line-forming regions.

At $+$42 days there is significant polarisation at the absorption minimum of the \halpha\,\,feature labelled ``notch C'' ($v_{\rm line}$$=-$7300 \kms) in the 1200R data. Its line polarisation was determined in the same way as that of the low-velocity \halpha\,\,feature discussed above. It is observed with $p_{\rm line}$=1.0$\pm$0.4\% and $\theta_{\rm line}$=141$\pm$12$^{\circ}$. There is no significant polarisation associated with other lines that could be interpreted as arising from the same line-forming region as this feature at this epoch.

In the later epochs of $+$64 and $+$73 days there are several polarised peaks in the range of $p_{\rm line}\sim$0.6-1.4\% for \halpha, \hbeta\,\,and He\,{\sc i} $\lambda 5876$ that have no obvious corresponding feature associated with other lines. The polarisation angles of these features mostly cluster around the same region of the sky at $\sim$120$^{\circ}$. At $+$64 days, 3 features are detected in the 300V data associated with \halpha\,\,at $-$7000 - $-$12150 \kms with polarisation at the 0.6-0.8\% level. In contrast, no significant polarisation is detected in this velocity range in the 1200R spectrum of four days later at $+$68 days. It is possible that these features have dissipated in that time, however  Figure \ref{1200rpolflux} shows that there are some rises in the polarisation in this range. The uncertainty on the continuum polarisation is higher at $+$68 days than in the 300V spectrum of $+$64 days; the larger uncertainty could mean that these features are present but not detected as being significantly different from the continuum polarisation at $+$68 days.

The Ca\,{\sc ii} IR triplet only becomes polarised at $+$64 days in the 300V data (see Figure \ref{300vpolflux}), associated with the strengthening of the emission lines at this epoch (Figure \ref{lineid}). Strong line polarisation is observed in the two distinct absorption troughs. The velocities of the polarised features were determined by using the nearest redward emission as the rest wavelength. Note that these may be inaccurate due to ambiguity in which of the components of the triplet is responsible for the feature. At $+$64 days the maximum observed polarisation associated with the Ca\,{\sc ii} IR triplet is $p_{\rm line}$=1.1$\pm$0.4\%, the components in both absorption troughs have similar consistent polarisation angles of $\sim$100$^{\circ}$. At $+$73 days both of these features strengthen. The feature in the trough of the 8542 \ang line (at $\lambda = 8375$ \ang) is most strongly polarised at $p_{\rm line}$=3.4$\pm$0.9\%, the polarisation of the $8662$ \ang line, on the other hand, increases to 1.7$\pm$0.9\%. It is probable that the blended nature of the emission has diluted the polarisation from the 8662 \ang line, such that it is not as highly polarised as the 8542 \ang line.  

\subsection{Evolution of SN 2009ip on the polar diagram}
\label{sec:polarplot}
 \begin{figure*}
\includegraphics[scale=0.23]{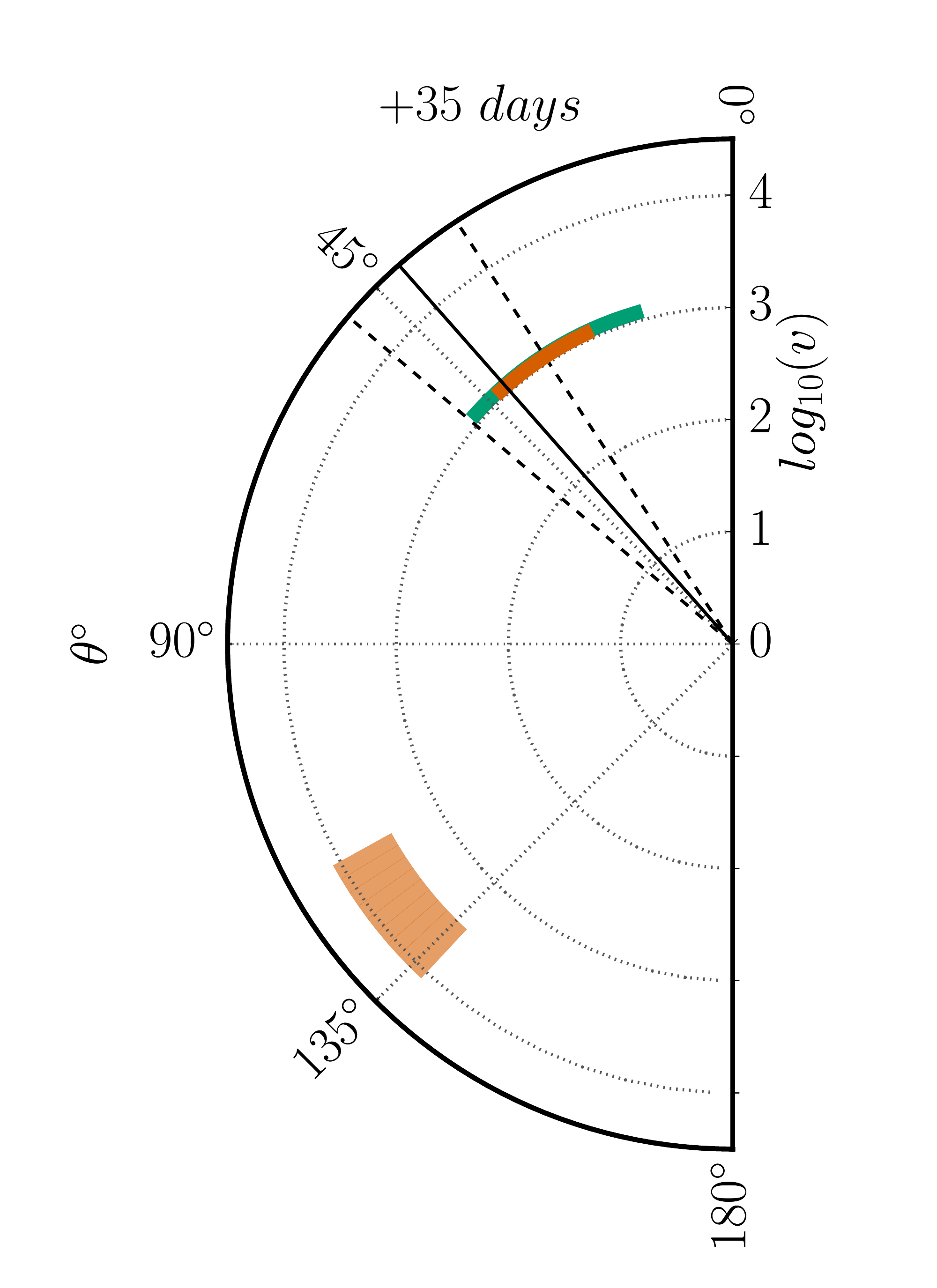}\includegraphics[scale=0.23]{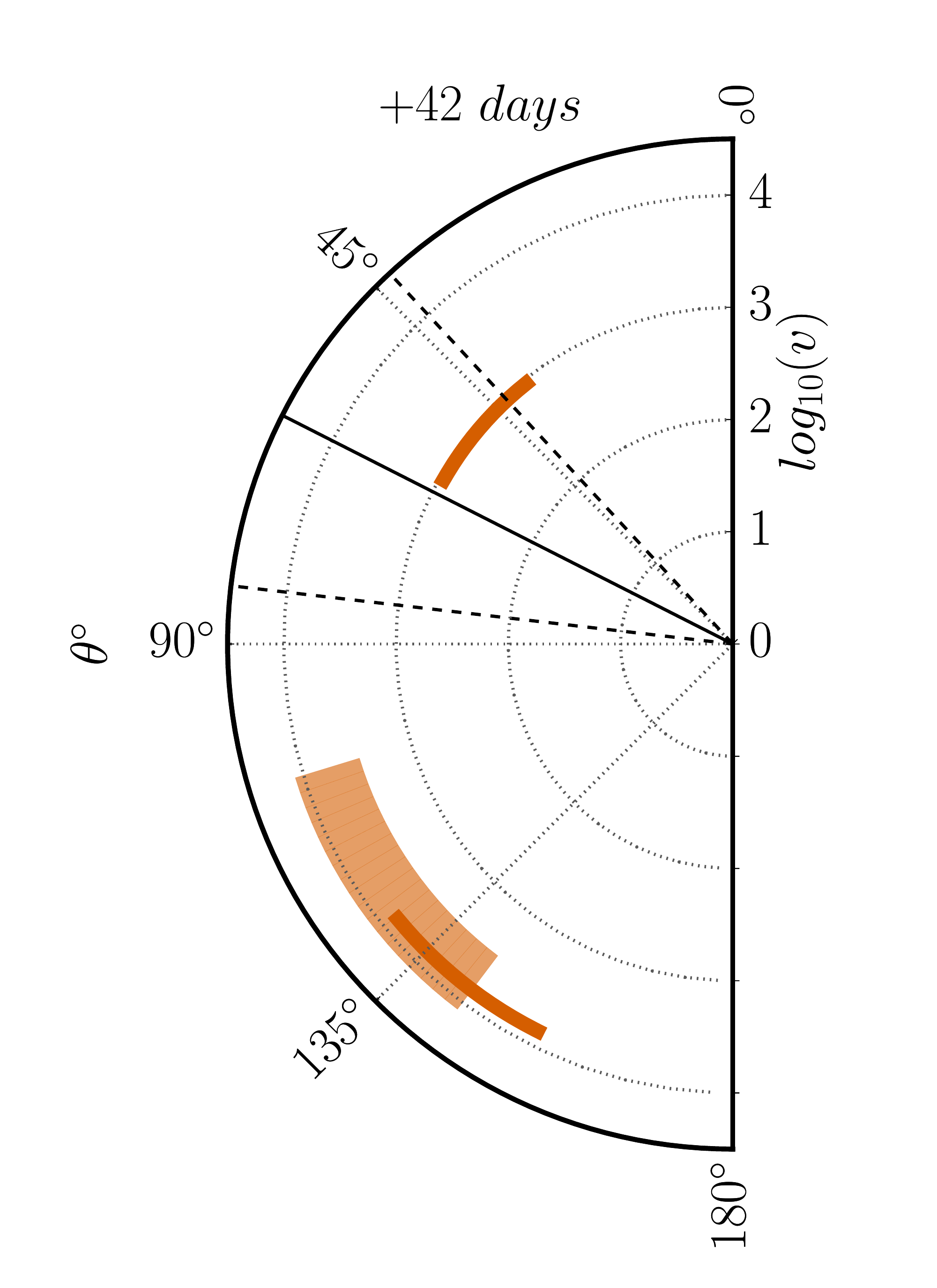}\includegraphics[scale=0.23]{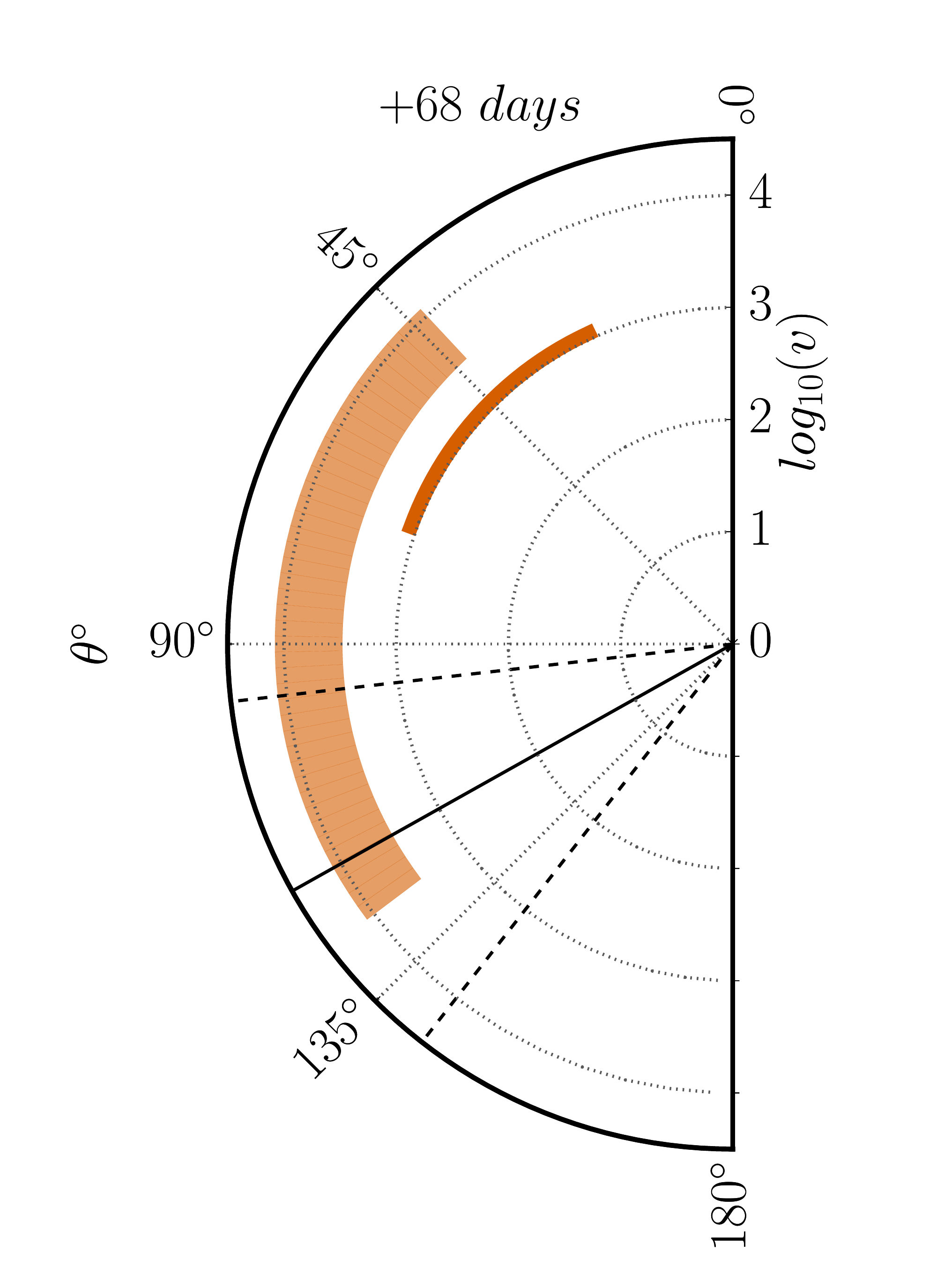}\includegraphics[scale=0.35]{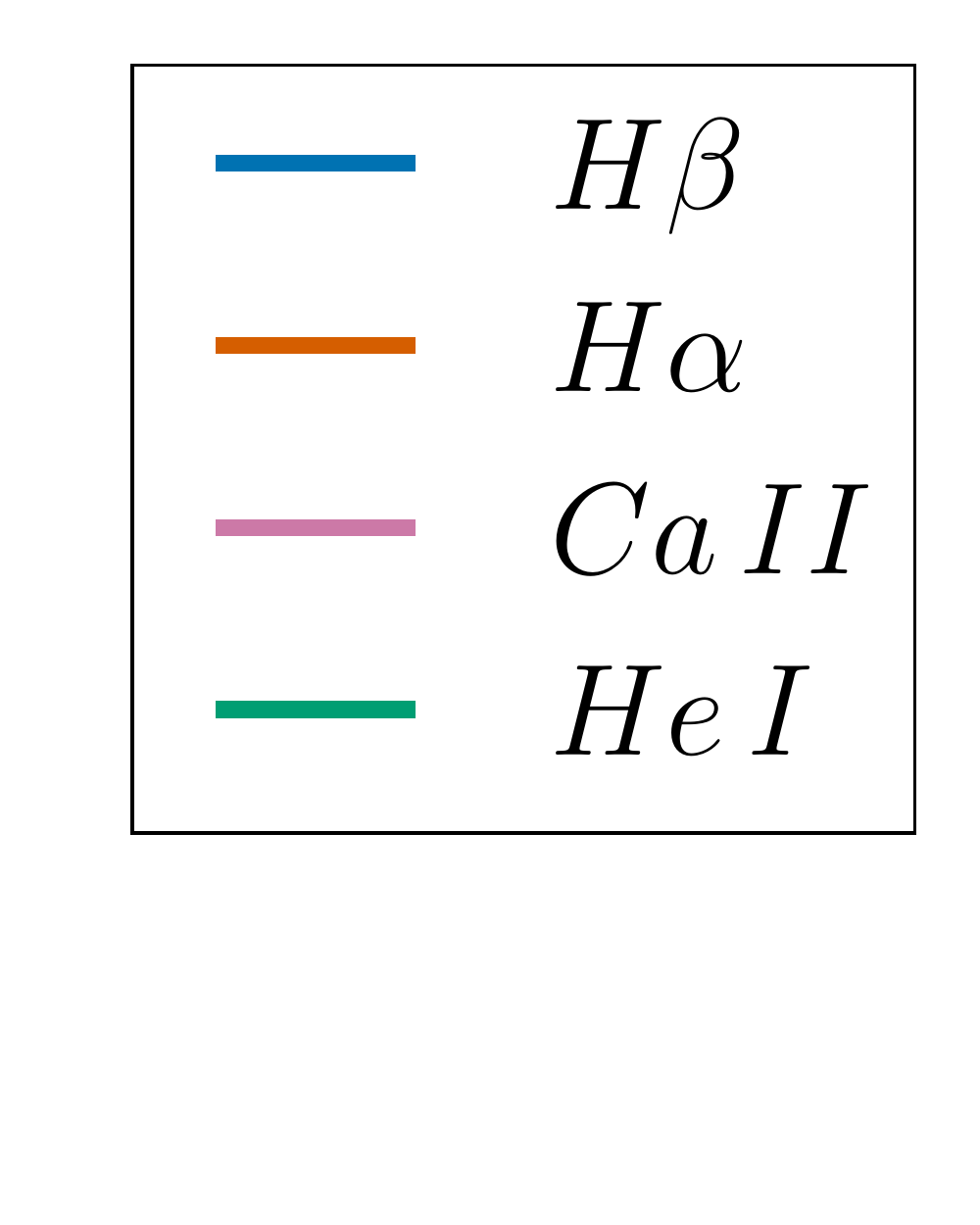}
\includegraphics[scale=0.23]{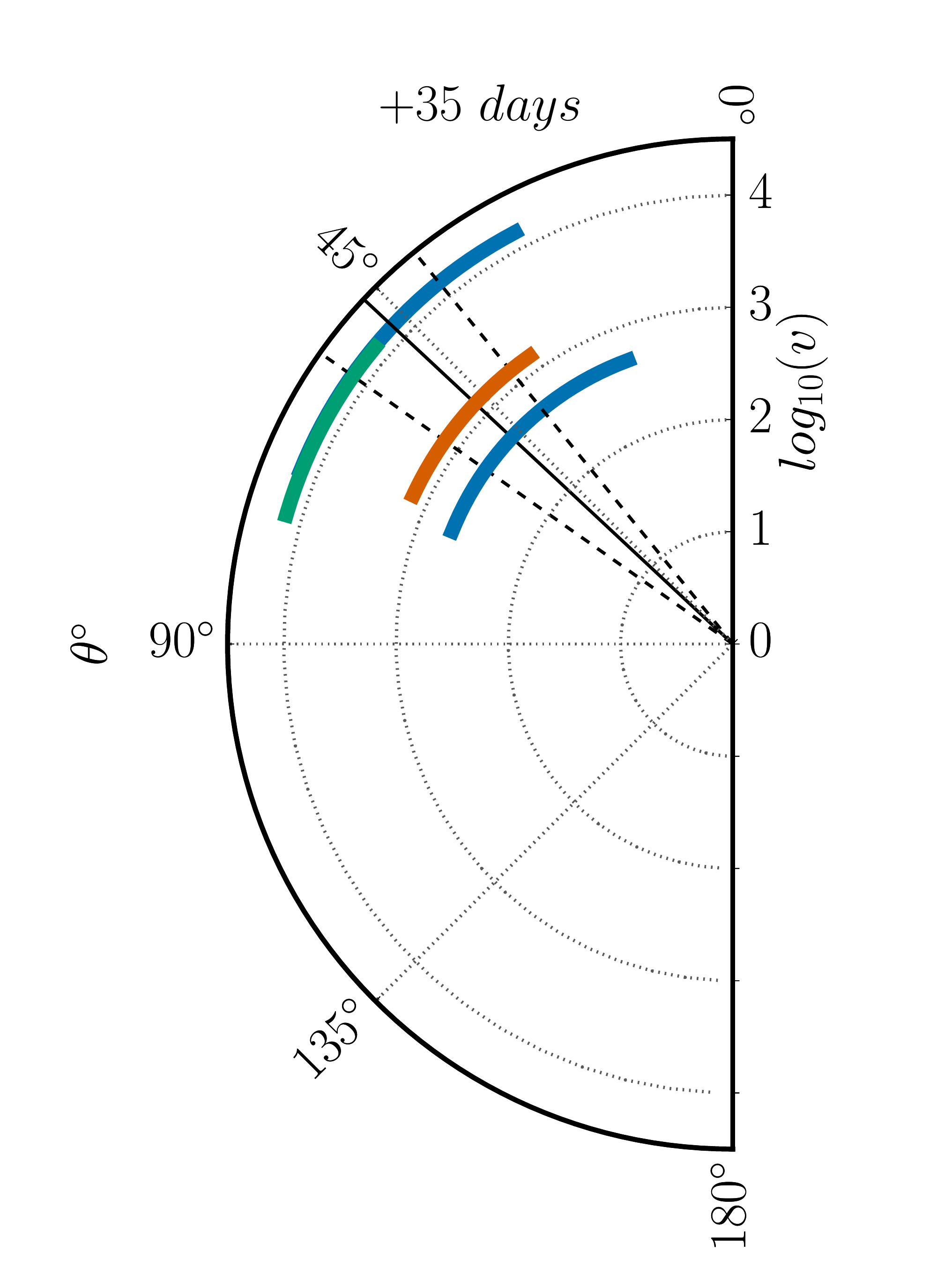}\includegraphics[scale=0.23]{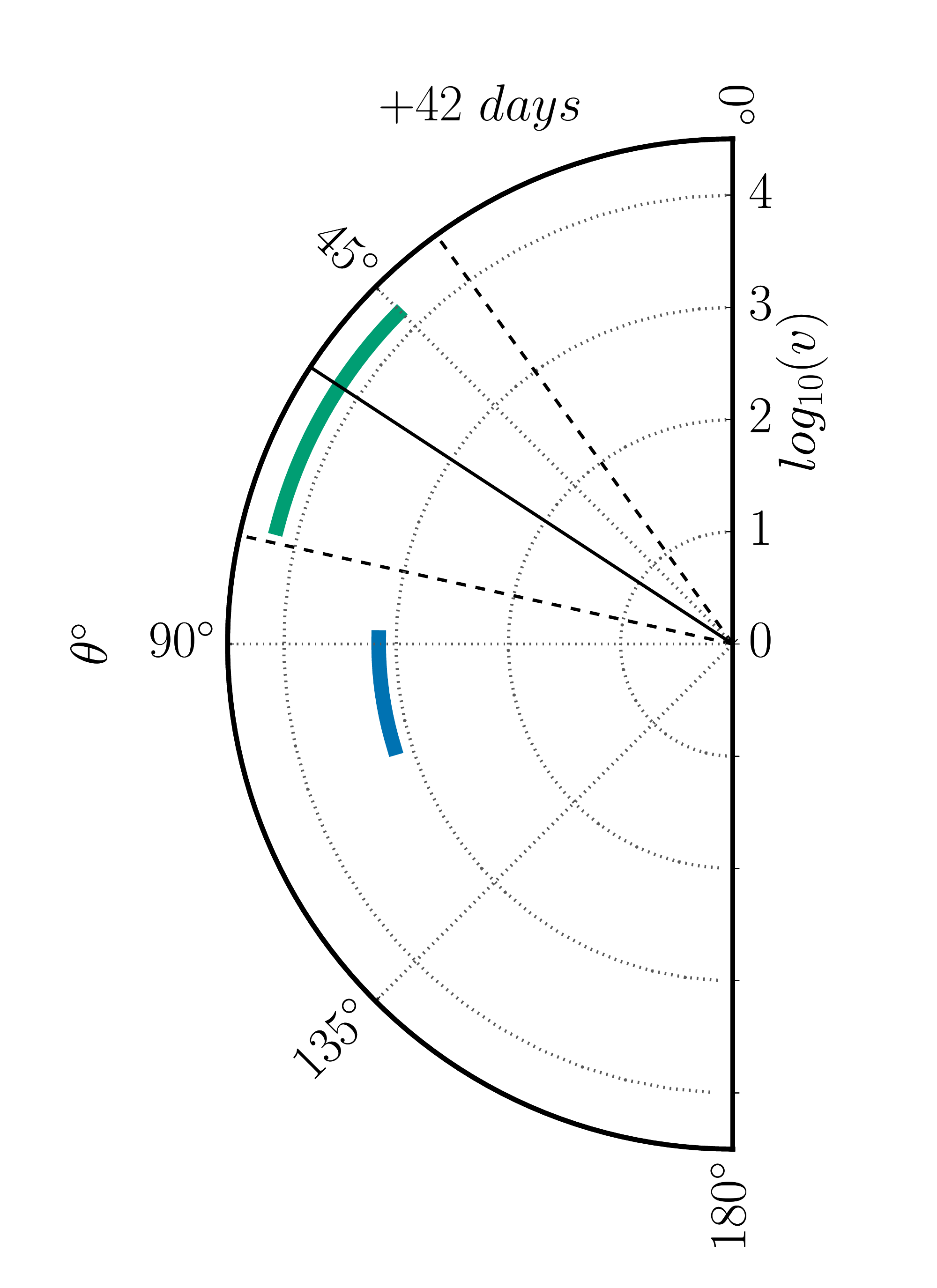}\includegraphics[scale=0.23]{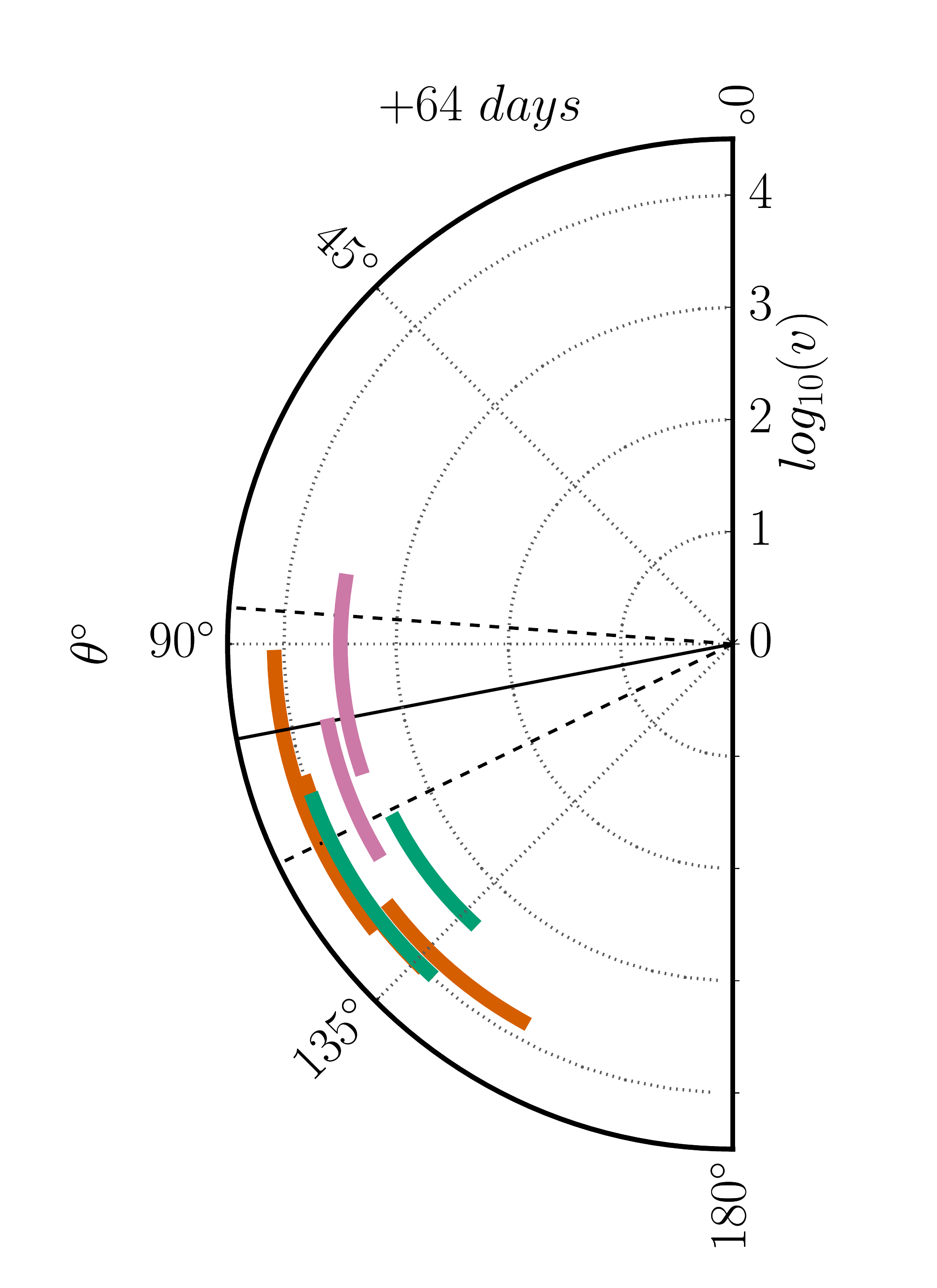}\includegraphics[scale=0.23]{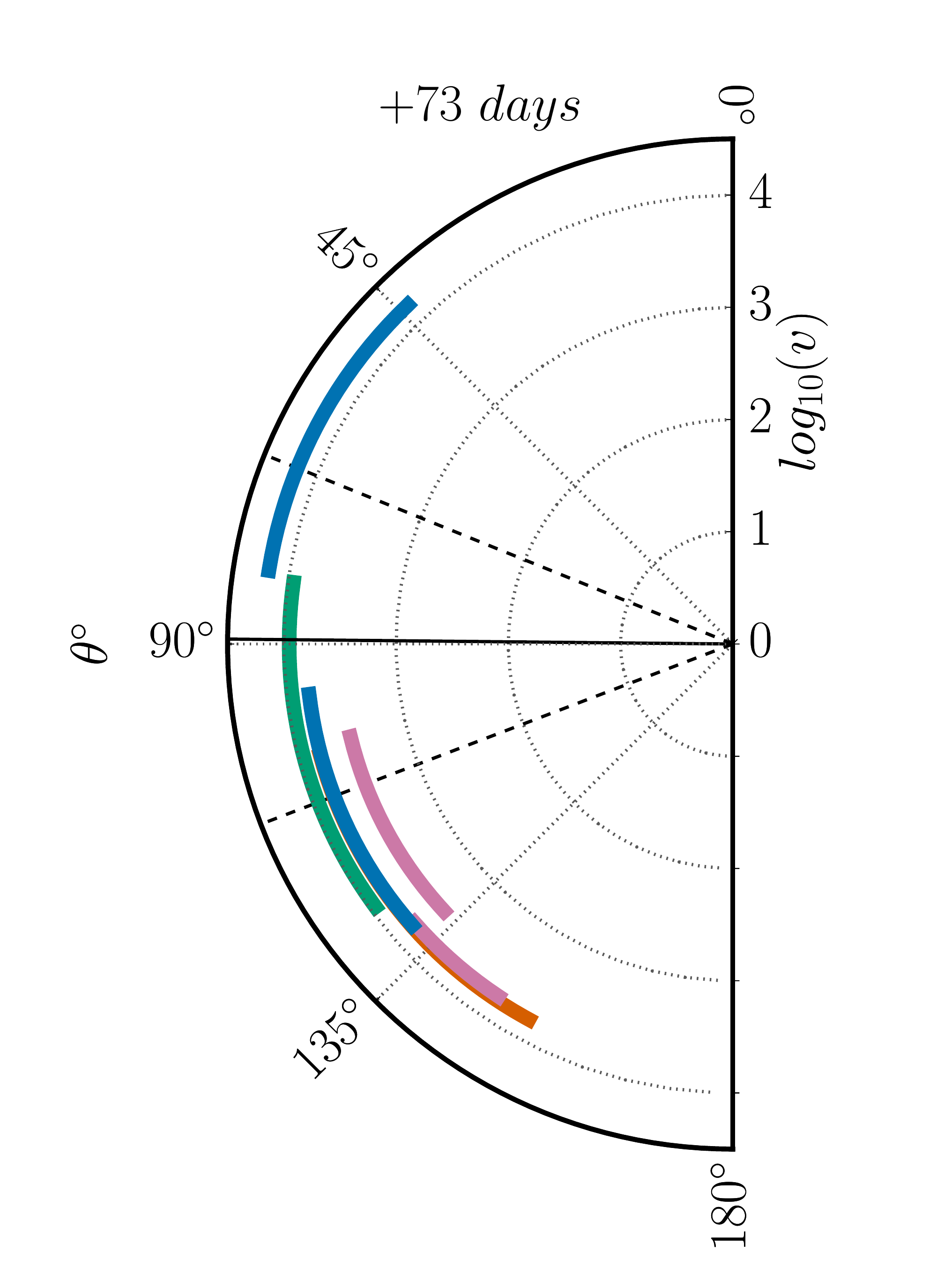}
\caption{Top panel; 1200R Polar plots of the polarisation angle of the \halpha\,\,and He\,{\sc i} $\lambda5876$ absorption lines, the velocity in $\log_{10}( v\, \mathrm{km\,s^{-1}})$ is indicated radially. Bottom panel; 300V Polar plots showing  the angle of \halpha, \hbeta, He\,{\sc i} $\lambda5876$ and Ca\,{\sc ii} IR triplet absorption polarisation.  The polarisation angle of the broad \halpha\,\,P Cygni absorption is shown by the slightly transparent orange wedge. The radial extent of the wedge represents that the measurement is the average of the polarisation angle between $-$12000 - $-$3000 \kms. The continuum is marked with the black lines with the 1$\sigma$ uncertainty indicated by the dashed lines.} 
\label{polarplots}
\end{figure*}

Figure \ref{polarplots} shows the evolution of the polarisation angle for different lines as a function of velocity. The polarisation angles of each species, as described above, are presented as arcs centred on $\theta_{\rm line}$, and the length of the arc represents the $\pm$1$\sigma$ uncertainty.

The 1200R data is particularly useful for studying the absorption component of the low-velocity narrow \halpha\,\,P Cygni profile. The polarisation angle of this feature does not display much temporal evolution, lying at approximately 45$^{\circ}$ during all of the observations. It is closely aligned with the low-velocity He\,{\sc i} feature and the continuum polarisation angle at $+$35 and $+$42 days. The broad \halpha\,\,P Cygni absorption line is polarised at $\sim$90$^{\circ}$ to the narrow feature. The high-velocity narrow notch labelled C ($+$42 days) is highly misaligned with the narrow P Cygni profile and the continuum at this epoch; instead it is aligned with the broad \halpha\,\,P Cygni absorption component.

The 300V data show that the polarisation angles of all species and the continuum at $+$35 days are concentrated around 45$^{\circ}$, including hydrogen and helium traveling at $\sim$$-$14000 \kms. At $+$42 days the continuum undergoes a slight counterclockwise rotation, which is observed in both the 300V and 1200R data. The low-velocity \hbeta\,\,polarised feature rotates to $\sim$90$^{\circ}$, where the high-velocity helium polarisation remains at 60$^{\circ}$. At $+$64 days the continuum polarisation angle has rotated further to $\sim$100$^{\circ}$, the material with $v>$4000 \kms is also clustered around the same position, between 90-150$^{\circ}$. This remains the case at $+$73 days except for one component of \hbeta\,\,at $-$15250 \kms that is polarised at $\sim$60$^{\circ}$. The continuum polarisation shows a slight clockwise rotation at $+$73 days, whereas its previous evolution was in a counterclockwise fashion. As mentioned in Section \ref{subsubsec:cont}, the evolution of the continuum polarisation angle is sensitive to the choice of the ISP.


\section{Discussion}
\label{sec:discussion}

\subsection{The geometry of interacting transients}
\label{dis:context}

The polarimetric observations of SN 2009ip add to the small sample of only 3 previous interacting transients studied with this technique, SN 1997eg, SN 1998S and SN 2010jl \citep{Hoffman2008, Leo2000, Wan2001, Pat2011, Bau2012}. All three SNe showed significant continuum polarisation in the $\sim$2-3\% range. Developing an interpretation of the geometry of Type IIn SNe can be a difficult task considering the ambiguity surrounding the question of what fraction of the observed polarisation is due to electron-scattering in an asymmetric photosphere as opposed to asymmetries in the interaction region. Given that SN 1997eg, SN 1998S and SN 2010jl were interaction-dominated transients, the high degrees of continuum polarisation have in some cases been interpreted as indicating significant departures from spherical symmetry in the circumstellar material. For these objects, the geometry of the CSM has been inferred to be disk-like. At late times ($>$100 days) both SN 1997eg and SN 1998S display broad double or triple peaked Balmer lines, with equatorial disk-like enhancement considered as the explanation. The elevated polarisation in the blue wings of \halpha\,\,of SN 1997eg led \citeauthor{Hoffman2008} to the conclusion that the CSM was disk-like for that SN. \citet{And2011} found that a disk or torus like geometry was required for SN 2010jl in order to explain the low optical depth along the line of sight in conjunction with the high IR excess observed. X-ray observations also indicate an aspherical CSM formation for SN 2010jl \citep{Cha2015}.

Despite the small sample size a somewhat consistent picture of these objects is emerging. The highly aspherical, toroidal or disk-like CSM observed in other Type IIn SNe is also invoked by \citet{Lev2014}, \citet{Fra2013}, \citet{Mauer2013}, \citet{Ofek2013} and \citet{SMP14} to explain various observations of SN 2009ip. \citeauthor{Lev2014} used this geometry to explain the observed low ratio of I(\halpha)/I(\hbeta), which required an extraordinarily high density for the CSM. \citet{Ofek2013} found that the mass loss rate derived from the \halpha\,\,luminosity was an order of magnitude higher than the mass loss derived from the X-ray and bolometric luminosity. This required that the narrow \halpha\,\,component be formed at radii larger than the shock radius in a spherical CSM or that the surrounding material was aspherical in geometry. It is apparent that an aspherical structure surrounding SN 2009ip is the natural explanation for many of the otherwise difficult-to-reconcile observations. 

Polarimetric observations of SN 2009ip at around peak maximum light also indicate that the CSM is highly aspherical. \citet{Mauer2014} published multi-epoch spectropolarimetric data from the 21$^{\mathrm{st}}$ September 2012 (during the 2012a re-brightening) to 10$^{\mathrm{th}}$ December 2012 ($+$68 days), and included data that is presented here. The V-band polarisation measurements made by \citet{Mauer2014} are shown in Figure \ref{continuumvtime} for reference. \citeauthor{Mauer2014} noted a rotation in the continuum polarisation angle of 90$^{\circ}$ degree between the 2012a observations (taken at $\sim$$-$12 days with respect to the 3$^{\mathrm{rd}}$ October 2012b maximum) and observations taken around the peak of the 2012b explosion (at $\sim+$11 days), from which they inferred orthogonal axes of symmetry for the structures of the 2012a and 2012b events. They suggested that the polarisation angle of the continuum during the 2012a event ($\sim$166$^{\circ}$) represented the axis of symmetry of the ejecta from a SN explosion occurring on 24${\rm th}$ July 2012 and that the onset of 2012b marked the beginning of interaction of that ejecta with an orthogonal disk-like circumstellar component. The orientation of the latter was inferred from the continuum polarisation angle of $\sim$72$^{\circ}$ measured around the 2012b peak.

The observations presented here are discussed in the following subsections within the context of the previous observations of SN 2009ip and interacting transients in general.

\subsection{Origin of the broad \halpha\,\,profile and continuum polarisation}
\label{dis:broadhalpha}
In Type IIn SNe, the broad component of \halpha\,\,may either arise in the reprocessing of photons due to electron-scattering in the shocked dense CSM \citep{Chu2001} or in the reverse-shocked ejecta. In the first case, electron-scattering would produce broad Lorentzian wings with photons up-scattered to high velocities through multiple scattering events. If the line-forming region is asymmetric, the up-scattered photons in the outer wings of \halpha\,\,would be significantly more polarised than the core. The increased polarisation is because the photons making up the outer wings are more likely to have undergone polarising scattering incidents (in order to reach the higher velocities) than those arising in the core of the line, some of which may have escaped without electron scattering. 

In Section \ref{subsubsec:broadhalpha} we analysed the polarisation signal of the red wing of the broad P Cygni component of \halpha. At both $+$35 and $+$68 days, there is little evidence to support an increase in the line polarisation in the outer wings compared to the core. This suggests that there is minimal reprocessing of \halpha\,\,by a shell of electrons and therefore makes formation of this component in an external optically thick CSM unlikely.  

The lack of reprocessing is not unexpected as the broad emission at this time dramatically differs from the Lorentzian profile observed at the peak of the lightcurve, instead it resembles a P Cygni profile (see Figure \ref{halphahbeta1200R} and \citealt{Mauer2013, Fra2013, Mar2014, Mauer2014}). The observed $\sim$90$^{\circ}$ separation in polarisation angle between the narrow and broad \halpha\,\,absorption lines (Figure \ref{polarplots}) would also indicate that the two components do not arise in the same line-forming region. 
 
Around the time of our first observation the continuum polarisation angle is observed to rotate significantly from earlier measurements (see Figure \ref{continuumvtime} and \citealt{Mauer2014}). The evidence suggests that the interaction had begun to fade before our observations (when the ejecta travelling at $\sim$8000 \kms would be $\sim6.7\times10^{10}$ km from the centre of the explosion) and that the broad lines and photosphere now reside in the ejecta rather than the interaction as is thought to be the case at maximum light.

Note that the beginning of our observations is concurrent with a ``bump" in the light curve at $\sim+$35 days (Figure \ref{continuumvtime}). There is no consensus on the source of this fluctuation (or the others present in the lightcurve). Suggestions for the origin of the bumps include heating from a surviving progenitor \citep{Martin2015}, interactions of shells ejected from a binary system \citep{Sok2013, Kas2013}, the temporary re-brightening of the ejecta photosphere \citep{Mauer2014} or interaction with an inhomogenous CSM \citep{Gra2014}. \citet{Martin2015} proposed that they could arise from the reheating of the inner rim of the disk-like CSM that \citet{Lev2014} suggested was responsible for the 2012b outburst. See \citet{Martin2015} for an in-depth analysis of the fluctuations in the light curve. The observations presented above suggest that, regardless of the source of the ``bump", the continuum polarisation is no longer dominated by electron-scattering in the interaction at the time of our observations.

\subsection{Origin of the low-velocity \halpha\,\,polarisation}

The observed polarisation of the low-velocity ($\sim-$1000 \kms) \halpha\,\,absorption may be used qualitatively to constrain the structure of the circumstellar medium. The relatively weak absorption strength coupled with the significant polarisation indicates that the CSM obscures a small and highly aspherical region of the photosphere from $+$35 to $+$68 days. This may point to the equatorial enhancement or disk-like structure surrounding the progenitor that was suggested in previous studies \citep{Fra2013, Mauer2013, Ofek2013, Lev2014, SMP14, Mauer2014}. The observed absorption depth and polarimetric signatures of such a CSM are dependent upon a number of parameters, such as the orientation of the plane of the non-spherical CSM, the extent perpendicular to that plane $z$ (the thickness of the disk), and the inclination with respect to the line of sight $i$. These parameters may then be used to rule out various configurations. For example, at certain inclinations a disk-like geometry may cover large portions of the photosphere and induce very strong absorption and polarisation if the photosphere is not uniformly obscured, at larger inclinations, however, it may not obscure the photosphere at all.

In an effort to understand the geometrical configurations that could produce the low-velocity \halpha\,\,polarisation, a model of the photosphere and line-forming region was constructed following the prescription of \citet{Wan2007} and \citet{Mau2010}, and described in detail in \citet{Rei2016}. The model simulated $1\times10^{7}$ photons across a limb-darkened 2D elliptical photosphere and tested the polarisation induced by blocking it with simple line-forming regions. To simulate the polarisation of the 2D photosphere, each photon was assigned either a random polarisation angle or one tangential to the ellipse at the position of the photon. This depended on the distance of the photon's origin from the centre of the photosphere, with photons arising closer to the limb less likely to be randomly polarised. The probability of non-random polarisation was set such that for a given axis ratio, the resulting continuum polarisation reproduced the polarisation expected for an oblate ellipsoid as specified by the models of \citet{Hof1991}. 

The polarisation observed in the 1200R data at $+$35 and $+$68 days was simulated. As the narrow \halpha\,\,absorption is strongly polarised but weakly absorbing, line-forming regions that would induce a strong polarisation while only absorbing a small fraction of the photospheric photons were chosen. These were assumed to completely absorb all photons originating from the area of the photosphere obscured by the line-forming region. The total normalised Stokes vectors of the escaping photons were calculated. The continuum, line and total observed polarisations were calculated in the same manner as for the observational data here. The fraction of photons absorbed by the line-forming region was calculated and compared to that absorbed in one 5 \ang\,\,bin in the observed 1200R flux spectrum. 

The observed continuum polarisation was reproduced at $+$35 and $+$68 days with an axis ratio of 0.88 (0.90 at $+$68 days), as expected from \citet{Hof1991}, implying deviations from spherical symmetry on the 10-15\% level for the photosphere for these epochs. Rotating the ellipse such that the long axis of the photosphere was at 135$^{\circ}$ (30$^{\circ}$ for $+$68 days) from the y axis (North) reproduced the continuum polarisation angle of 41$\pm$8$^{\circ}$ (119$\pm$22$^{\circ}$ for $+$68 days). The line-forming regions were first tested for the data at $+$35 days before successful models were then tested again given the continuum polarisation at $+$68 days. Two line-forming regions were found to successfully approximate the line and total observed polarisation and the absorption depth of the narrow \halpha\,\,profile at $+$35 days. These represent two possible scenarios for the circumstellar medium: a disk-like structure observed edge-on with the photosphere oblate with respect to the plane of the disk or alternatively a similar structure inclined with respect to the observerand the photosphere prolate to the plane of the disk. These scenarios are shown in Figure \ref{fig:schematic} and the observed and simulated data can be seen in Table \ref{table:model} in the Appendix.

\subsubsection{Edge-on disk-like CSM scenario}
\begin{figure}
\includegraphics[scale=0.5]{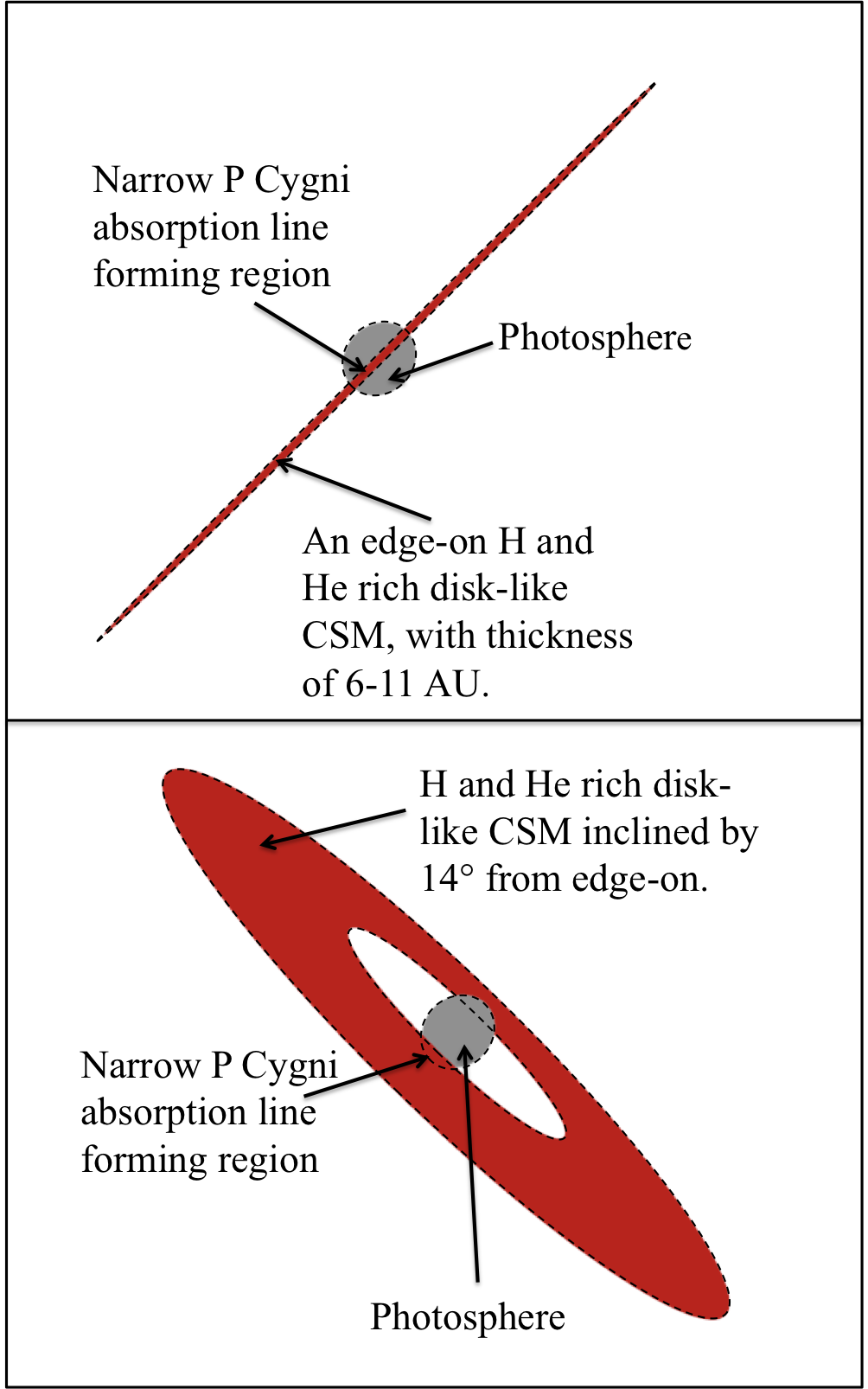}
\caption[Schematics of the geometries that with the Monte Carlo simulation successfully reproduced the polarisation of the continuum and the low-velocity \halpha\,\,feature of SN 2009ip at $+$35 days.]{Schematics of the successful geometries that reproduced the polarisation of the continuum and the low-velocity \halpha\,\,feature at $+$35 days. The top panel shows the scenario where the observer views a disk-like CSM edge-on. In this case the photosphere is oblate with respect to the long axis of the CSM. The bottom panel shows the scenario where this geometry is inclined out of the line of sight by 14$^{\circ}$, where the photosphere appears to be prolate with respect to the long axis of the CSM.} 
\label{fig:schematic}
\end{figure}

A thin rectangular line-forming region, similar to that projected by an edge-on disk-like or equatorial enhancement, was found to approximately reproduce the observed properties of the low-velocity \halpha\,\,feature. The line-forming region, as shown in the top panel of Figure \ref{fig:schematic}, had a long axis that aligned with that of the photosphere. The absorption depth and degree of polarisation were found to be sensitive to the length of the short axis (corresponding to the extent perpendicular to the plane of the CSM or the thickness of the disk). In order to reproduce the polarimetric properties, a short axis with an extent of 11$\pm$1\% that of the long axis of the photosphere was required. This corresponded to 11$\pm$1 {\sc au} ($\sim 1.65 \times 10^{9}$ km) assuming a photospheric radius of $1.5 \times 10^{10}$ km at $+$35 days \citep{Mar2014}. This was found to overestimate the absorption depth by $\sim$10\%. A thinner line-forming region of 6 {\sc au}  ($\sim 0.90 \times 10^{9}$ km) could reproduce the fraction of absorbed photons but was inconsistent with the observed degree of polarisation. The first configuration was also found to successfully reproduce the polarimetry at $+$68 days, the absorption depth however was larger than that observed at $+$35 days, which is inconsistent with the observations, considering that the low-velocity feature is not observed in absorption in the latter epoch (Table \ref{table:model}).

The orientation of this CSM parallel to the major axis of the photosphere has the implication that the continuum polarisation at $+$35 days arises in a pseudo-photosphere created by electron-scattering at the interface between the CSM and the ejecta. Furthermore, the disappearance of the narrow \halpha\,\,absorption and the polarisation at $\sim+$70 days imply that the ejecta have swept up or passed the outer edge of the CSM by this epoch. Considering an explosion epoch of 24${\rm th}$ July 2012 (the onset of 2012a) and a velocity of $-$8000 \kms\,\,for the bulk of the ejecta, the outer edge of the disk can then be constrained to $10.2 \times 10^{10}$ km.

\subsubsection{Inclined disk-like CSM scenario}
\label{dis:inclineddisk}
The observed properties were also reproduced by absorbing photons in the limb of the long axis of the photosphere (as seen in the bottom panel of Figure \ref{fig:schematic}); similar to how a disk-like CSM that is slightly inclined with respect to the observer would appear. In this case the photosphere is prolate with respect to the plane of the CSM. This was performed by taking a disk-like CSM with a circular inner radius and inclining it incrementally. Assuming a bulk velocity flow of $-$8000 \kms\,\,for the ejecta, an explosion epoch corresponding to the onset of 2012a (24$^{\rm th}$ July 2012) and the beginning of 2012b (23$^{\rm rd}$ September 2012) as marking the onset of the interaction of the ejecta with the CSM, the position of the inner edge of the CSM was estimated to be $4.2 \times 10^{10}$ km. Inclining this structure by 14$\pm$2$^{\circ}$ (given a photospheric radius of $1.5 \times 10^{10}$ km, \citealt{Mar2014}) with respect to the observer approximately simulated the observed polarisation at both $+$35 and $+$68 days. The absorption depth was also slightly closer to that observed at $+$35 days and showed a larger decrease for $+$68 days, which is more consistent with the observations than that simulated with the edge-on model CSM above. The simulated Stokes para-meters were sensitive to changes of a few degrees in the inclination, however uncertainty in the position of the inner radius of the disk results in an increase in the uncertainty of the inferred inclination of the CSM disk. 

By comparing the black body radius inferred by \citet{Mar2014} to the projected extent of the inner radius of the simulated CSM, we determined that the disk would begin to block the photosphere and produce an absorption feature just prior to the 2012b maximum. The absorption feature would then disappear at $\sim$$+$70 days as the photosphere retreated. Indeed, the polarised feature is observed to disappear between $+$68 and $+$83 days. In this scenario the low-velocity \halpha\,\,polarisation is produced by the same material that interacts with the ejecta at earlier times to power the 2012b maximum.\\

\subsubsection{A model geometry for the CSM}
The alignment of the photosphere with the plane of the CSM in the edge-on disk scenario at $+$35 days may imply that the continuum photons are polarised by electron-scattering in the interaction region (a pseudo-photosphere). This is difficult to reconcile with the inferred lack of reprocessing of photons discussed in Section \ref{dis:broadhalpha}. Similarly, the orientation of such a pseudo-photosphere in these observations at $\sim$40$^{\circ}$ is inconsistent with the continuum polarisation angle observed during the 2012b maximum of $\sim$70$^{\circ}$ (\citealt{Mauer2014}; their $V$-band polarisation observations are shown in Figure \ref{continuumvtime} for reference). \citet{Mauer2014} suggested that the continuum at that time arose in the interaction between a disk-like CSM and the ejecta. In the inclined-disk scenario, the production of the continuum polarisation in a pseudo-photosphere is unnecessary as the photosphere lies in a plane perpendicular to that of the CSM. This suggests that the photosphere instead arises in the ejecta as is also implied by the lack of reprocessed photons at these epochs. It is also difficult to understand the presence of an edge-on disk-like CSM out to $10.2\times10^{10}$ km, given that the transient had become ejecta-dominated before our observations, implying an outer radius of $\sim6.7\times10^{10}$ km for the CSM.

Predictions of the emergence and disappearance of the absorption in the inclined-disk scenario match the observations and explain the lack of narrow absorption during the 2012a outburst. Under the assumption that the observed CSM was formed before the explosion, an edge-on disk would be expected to produce an absorption during the early outburst which is not observed. This discrepancy may be avoided under certain formation scenarios which suggest that the observed geometry was created during or after the 2012a event. It is, however, difficult to avoid the special circumstances that would be required to observe a thin disk edge-on. While in the reference frame of the star such a structure is just as likely to form at any orientation with respect to the star, placing that structure along the line of sight to Earth would be statistically remarkable. 

The data at both $+$35 and $+$68 days were also slightly better reproduced with the inclined-disk model (Table \ref{table:model}). For the reasons above, we favour the inclined scenario simulated here and shown in the bottom panel of Figure \ref{fig:schematic}. Given the shared absorption and polarised components of hydrogen and helium at low-velocity (Figures \ref{halphahbeta300V}, \ref{halphahbeta1200R} and \ref{polarplots}), the helium feature likely shares this inclined-disk geometry.  

We note that the successful simulation of the data with two distinct geometries highlights that this simple model suffers from degeneracies and since there are two scenarios that replicate the data well, there may be more. Similarly, the scenarios determined here may not reproduce the polarimetry under models that include the density of the CSM and full radiative transfer.


\subsection{Polarimetric implications for the geometry}
From our spectropolarimetric data we have identified two distinct phases in the evolution of SN 2009ip: phase 1- the geometry inferred from the first epoch of our observations and phase 2- the geometry inferred at around  $\sim+$70 days. 

The first phase occurs at $+$35 days. At this epoch strong line polarisation reveals a hydrogen and helium rich circumstellar disk, or equatorial ring, inclined by $\sim$14$^{\circ}$, such that it blocks a portion of the photosphere. The continuum polarisation indicates that the ejecta was oriented prolate with respect to the plane of the disk at this time (bottom panel of Figure \ref{fig:schematic}).  Using the models of \citet{Hof1991}, the continuum polarisation of $\sim$0.8\% at $+$35 days implies an axial ratio of $\sim$0.85 (or 15\% asymmetry) for an ellipsoidal photosphere and indicates a bipolar structure for the SN ejecta. We observe that the polarised components of high-velocity He\,{\sc i} $\lambda 5876$ and \hbeta\,\,align with those at low velocity (Figure \ref{polarplots}), indicating that this material obscures the region of the photosphere as the CSM disk. 

In Section \ref{dis:broadhalpha} we discussed that the change in the \halpha\,\,line profile, in conjunction with the rotation of the continuum polarisation at around $+$35 days compared to earlier observations, signified that the outburst was no longer interaction-dominated. The persistence of narrow lines at these epochs implies that interaction is ongoing but does not dominate the continuum light. Viewing the ejecta while there is ongoing interaction is possible in the light of our polarimetric results, which favour a highly-aspherical inclined CSM. This would allow the ejecta to expand freely along the axis orthogonal to the CSM, while the CSM would also not block the entirety of the photosphere from the observer. This scheme has previously been suggested by \citet{SMP14} and \citet{Fra2013} to explain the high velocities observed in SN 2009ip and by \citet{Mauer2014} to explain their observation of polarised He\,{\sc i}/Na\,{\sc i}\,{\sc d} that was blue shifted with respect to the absorption minimum in the flux spectrum.

The orthogonality of the broad and narrow P Cygni absorption of \halpha\,\,(Figure \ref{polarplots}) may indicate that the fast-moving ejecta lie above and below the equatorial disk. In this way, the presence of the disk prevents it from blocking the same region of the photosphere that the disk obscures. The observed polarisation, therefore, does not necessarily imply an intrinsic deviation from spherical symmetry for the broad \halpha-forming region of the ejecta.The continuum polarisation at $+$35 days however, indicates that the photosphere is elongated with the long axis approximately perpendicular to the plane of the CSM, suggesting asymmetry in the fast-moving ejecta.

The aspherical ejecta could be the result of inherent asymmetry in the explosion mechanism (with the major axis orthogonal to that of previous eruptions which formed the CSM) or due to shaping of the ejecta by interaction with the surrounding medium. In the case of genuine deviation from spherical symmetry in the explosion mechanism, an axis of symmetry that is orthogonal to that of the CSM seems unlikely. Certainly, orthogonal components were not observed in Type IIn SN 1997eg, where, likewise, \citet{Hoffman2008} found a dual-axis of symmetry. The circumstellar and ejecta components in that case, however, were separated by 12$^{\circ}$ not 90$^{\circ}$ as is the case here. A more likely scenario for SN 2009ip is that the ejecta are forcibly elongated along the orthogonal axis due to interaction with the CSM, i.e. the circumstellar material prevents the ejecta expanding in that direction. This implies that the CSM is of a mass high enough to change the momentum of the ejecta.

The observations made in December 2012 ($\sim+$60 - $+$70 days) represent a new phase in the evolution of SN 2009ip, seen photometrically, spectroscopically and now observed in the polarimetric data as well. The rotation and decrease of the continuum polarisation between $+$35 days and $+$68 days (Figure \ref{continuumvtime}) suggests that the photosphere of the ejecta changes shape from a prolate ellipsoid with long axis at 45$^{\circ}$ to an oblate ellipsoid with the long axis at $\sim$120$^{\circ}$. Such a change in the geometry may be the result of a jet-like explosion \citep{Hof1999}. The change in the shape of the photosphere is coincident with a steep decline in the photospheric radius as determined by \citet{Mar2014}; indicating that in the later epochs the inner core of the explosion is revealed.

The continuum polarisation at $+$68 days of $\sim$0.5\% corresponds to an axial ratio of $\sim$0.9 or 10\% asymmetry, implying that the inner ejecta of the explosion are slightly more spherical than the outer envelope. The major axis of the photosphere rotates from the previous orientation at $+$35 days to approximately align with the axis of the disk-like CSM. Note that this rotation is partly dependent upon the choice of the ISP and that no physical rotation in the ejecta is implied, instead a rotation may result from a change in the underlying energy distribution. We suggest this is the result of interaction of the ejecta with the surrounding medium leading to aspherical cooling, and thus aspherical recombination of the outer envelope in the later epochs. 

The decreasing degree of polarisation at low velocities, as well as the almost complete disappearance of the narrow P Cygni absorption component in the later epochs, implies that the narrow line-forming region is much smaller in relation to the size of the photosphere than it was at $+$35 days. The simulation performed in Section \ref{dis:inclineddisk} suggests that this is the consequence of the rotation of the photosphere before $+$68 days, such that the inclined CSM blocks a much smaller region of the photosphere at this epoch. 

The polarisation angle for the narrow and broad absorption components of \halpha\,\,remains fixed from $+$35 days. In contrast, high-velocity He\,{\sc i} $\lambda 5876$, \hbeta, Ca\,{\sc ii} and the \halpha\,\,``narrow notches" generally evolve following the continuum polarisation angle. The flux spectra alone show signatures of the complex velocity structure of the ejecta observed in both the Balmer and He\,{\sc i} $\lambda 5876$ lines. The polarisation spectra reveal an asymmetric, shared and possibly clumpy line-forming region for these species.

\subsection{Origin of the CSM}

The observation of a disk-like structure for the CSM raises the question of its formation, whether preformed in an LBV wind, previous eruptions, through the interaction with a binary companion or, indeed, formed during the 2012 explosion itself. The formation of this disk in a cool LBV wind seems unlikely considering that the radial velocity is on the order of $\sim$1000 \kms\,\,which is much higher than the typical LBV terminal velocity of $\sim$200 \kms \citep{Dav2005}. Additionally, \citet{Dav2005} found evidence of clumpiness at the base of LBV winds rather than axisymmetry. It is possible, however, that the observed CSM was formed by an asymmetric explosion in one of SN 2009ip's many previous outbursts. The ejecta at $+$35 days, presuming that it was launched in July 2012, would have caught up to material that was launched in July-August 2011. During this period of time the putative LBV progenitor exhibited erratic outbursts \citep{Pas2013} and thus the observations here perhaps suggest a degree of asymmetry for these eruptions. 

A somewhat natural explanation for asymmetry in pre-existing CSM would be interaction with a binary companion. \citet{Sok2013}, \citet{Kas2013} and \citet{Tse2013} have discussed the outbursts of SN 2009ip in the context of binary interaction between an LBV and a more compact companion which is proposed to launch jets upon accretion from the more massive object. Such interactions in the pre-explosion system could lead to a bipolar arrangement for the CSM \citep{Mcl2014}. \citet{Smi2015lonelyLBV} suggested a binary rejuvenation or merger scenario for the origin of massive LBV stars in environments without large star-forming regions (although \citealt{Hum2016} have disputed this analysis). Such an environment for SN 2009ip was confirmed by \citet{Smi2016}, who also attribute the 2009-2011 outbursts to close binary encounters which could be the origin of the equatorial disk observed here. Binary interactions have also been suggested as the origin of the variability in $\eta$ Carinae (see, for example, \citealt{Sok2001} and \citealt{Smi2011}).

Alternatively, the inferred geometry may not be representative of the shape of the CSM surrounding the progenitor prior to the 2012 eruptions, but formed during the outburst itself. An asymmetric distribution of energy within a spherical construct produces an oblate structure in the plane orthogonal to that with the larger energy deposition \citep{Hof1999}. The scale height (or thickness) of the resulting toroidal shape is dependent upon the asymmetry of the explosion energy distribution. In this scenario, the observed orthogonality of the CSM and ejecta symmetry axes is a natural consequence of the expansion of the bipolar ejecta of the 2012a outburst into a spherical medium. The temporal evolution of the degree and angle of continuum polarisation as the ejecta interact with different portions of the CSM would depend on the resulting structure; the changing geometry of the electron-scattering surface with time and the optical depth \citep{Hof1995b, Hof1999}.

The disk-like structure may form part of an hourglass or bipolar nebulae as observed around $\eta$ Carinae, SN 1987A and commonly seen around LBV stars \citep{Wei2003, Smi2016etacaranalogue} of which we only observe the equatorial enhancement in the narrow absorption. \citet{Chi2008} present models that describe the formation of the hourglass structures observed around blue supergiants through a fast wind-slow wind interaction. They discuss a scenario in which a star at the onset of a blue loop phase at close-to-critical rotation ejects a dense equatorial disk \citep{Heg1998}. The following fast wind ejected by the blue supergiant then proceeds to sweep up the slower moving material into an hourglass structure. This follows from similar models performed by \citet{Lan1999} which describe the formation of the Homunculus nebula in the Great Eruption of $\eta$ Car. In that scenario, the LBV wind becomes slow and very dense during outburst, while the close-to-critical rotation results in the confinement of this wind close to the equator. An ensuing fast wind sweeps up the material into two thin-shelled lobes. \citet{Groh2009} suggested that the close-to-critical rotation required in these models to eject an equatorial disk is a typical attribute of strongly active LBVs.

An LBV progenitor to SN 2009ip raises the possibility of an $\eta$ Car-like CSM with bipolar lobes in addition to the equatorial enhancement observed here. In the inclined-disk scenario, it is unclear whether or not the ejecta are still interacting with the equatorial skirt. The narrow Balmer emission lines suggest interaction is ongoing. The lack of significant reprocessing of photons by electron-scattering indicates, however, that strong interaction with a dense CSM is no longer dominant. As the bipolar lobes of $\eta$ Car are much larger in extent than the equatorial disk, it is possible that the ejecta of SN 2009ip continue to interact with any CSM lobes throughout these observations. The observed narrow emission lines and possibly the ``bump" in the light curve may be produced this way. It is however, unlikely that the presence of any hourglass shape would block the photosphere in a way as to have an impact on the polarisation. The bipolar lobes would be much larger in extent than the photosphere and therefore would occlude it entirely, leading to the complete cancellation of the electric vectors of the light. It is possible that, similarly to $\eta$ Car, the progenitor to SN 2009ip was surrounded by a Homunculus-shaped nebula, however the polarimetric observations here reveal only an equatorial density enhancement.


\subsection{Physical constraints given the geometry}

SN 2009ip has been a well-documented transient, and numerous models have been proposed in an effort to understand the observations. Three categories of models have been proposed to explain the 2012 events; a terminal supernova explosion; another "impostor" outburst of an LBV star and ensuing interaction or through close encounters and a possible merger with a companion star. 

 Among the non-terminal eruption models is the pulsational pair instability eruption favoured by \citet{Pas2013} and \citet{Fra2013}, however the model offers no prediction for the geometry of the explosion or interaction. Thus the polarimetric data can not aid us in distinguishing it from other models of the outbursts. \citet{Mar2014} considered the energetics of the collision of the ejecta from the 2012b event with a spherical CSM. This required an explosion energy of only $\sim 10^{49} \mathrm{ergs}$, considerably lower than those required for a SN explosion. The disk-like geometry of the CSM inferred from the spectropolarimetry, however, implies that the explosion energy derived by \citeauthor{Mar2014} is likely to be significantly underestimated. A disk-like CSM would interact with a much smaller portion of the ejecta than a spherically symmetric CSM, therefore requiring much larger kinetic energy in the explosion in order to reproduce the observed luminosity. \citet{Mauer2014} estimated that only 2-3\% of the ejecta would intercept a disk of thickness $\sim$10 {\sc AU}, the extent of the CSM necessary to explain the line intensity of \halpha\,\,\citep{Lev2014}. Instead the observations may point to the 2012 events as being the result of a terminal SN explosion of energy $\gtrsim10^{51}$ ergs \citep{SMP14,Mauer2014}. This argument is only valid under the assumption that the geometry observed here is a structure that existed prior to the 2012a explosion and not, as the case may be, formed during the expansion of a highly-aspherical ejecta into a spherical CSM. In the latter scenario, the ejecta would encounter a much larger fractional area than it would with a pre-existing disk. 

The polarimetry is indicative of a bipolar explosion for the 2012 outbursts, consistent with the predictions of the non-terminal binary interaction models of \citet{Sok2013}, \citet{Kas2013} and \citet{Tse2013}. The predicted geometry of SN 2009ip also fits in quite well with the bipolar structure commonly observed in LBV nebulae \citep{Wei2003}. The existence of a disk (torus) around $\eta$ Car \citep{Mor1999} implies that such circumstellar media do exist around some LBV stars, \citet{Dav2005} found no evidence for this type of structure in the winds of 3 LBVs. \citet{Not1995} suggested that an equatorial density enhancement is required to create the bipolar nebulae surrounding LBVs. If SN 2009ip survived the 2012 outbursts, the observations of a probable disk (or torus) here, may be evidence of the proposed method of forming the bipolar structure. 

The continuum polarisation angle measured at the 2012b peak (72$^{\circ}$) is significantly different from the polarisation angle of the narrow \halpha\,\,component ($\sim$41$^{\circ}$), from which we determine the inclined-disk or torus structure for the CSM. The difference in the inferred orientations of the disk could indicate an evolving CSM shape, such as that expected from an aspherical explosion interacting with a spherical CSM. It does, however, seem unlikely that the orientation of the 2012a outburst ($\sim$166$^{\circ}$) would result in the CSM residing in the plane indicated by the low-velocity line polarisation observed here.  Alternatively, it is also possible that the continuum polarisation at the 2012b peak is the result of a pseudo-photosphere generated by the interaction of the ejecta with the inner edge of the same disk that we observe at $+$35 days (with no discrepancy). The polarisation that would arise from the interaction zones of either the pre-existing disk or the spherical CSM scenario is unclear without complex 3-D modelling of the interaction. 

We have shown that the continuum polarisation angle is subject to the choice of ISP and to temporal evolution, rotating between the inferred orientations (from line polarisation) of the ejecta and the CSM over time. The continuum polarisation at any one time may not be representative of the general orientation of the ejecta or of the interaction. We therefore suggest caution when interpreting the physical implications of the continuum polarisation in SN 2009ip and perhaps in all Type IIn SNe, where there may be multiple sources of continuum light. 

\subsection{SN 2009ip in the context of Core-Collapse Supernovae}
Spectroscopically, SN 2009ip has been shown to be similar to Type IIP SNe \citep{Gra2014, Mar2014, Mauer2013, SMP14} and indeed if the underlying explosion of SN 2009ip was a true core collapse event it is expected to resemble a Type IIP SNe. Here we compare the polarised properties of SN 2009ip to those of Type IIP, Type IIn and to the stripped core-collapse SNe in an effort to further classify SN 2009ip within these groups.

In general Type IIP SNe show moderate levels of continuum polarisation ($\sim$0.1-0.5\%) while in the plateau phase, this drastically increases to, in some cases, $\gtrsim$1\% when the photosphere recedes to reveal the inner asymmetry of the explosion mechanism \citep{Leo2006}. Stripped-core SNe have been shown to display relatively low levels of continuum polarisation ($\sim$0.4\%) but line polarisation levels of $\sim$4\%, particularly with the Ca\,{\sc ii} IR triplet (e.g. \citealt{Mau2009}). The maximum continuum polarisation observed with SN 2009ip is $\sim$1.7\% at around the peak of the 2012b light curve \citep{Mauer2014}. The geometry of the photosphere, however, is likely to be governed by the interaction at this stage and not purely by the ejecta. At $+$35 days the continuum polarisation, thought to arise in the fast-moving ejecta, is found to be $\sim$0.8\%. This is slightly higher than typically seen for Type IIP SNe during the plateau phase and is more similar to the levels displayed by SN 2004dj and SN 2007aa around the end of the plateau phase \citep{Leo2006, Chor2010}. This polarisation, however, is also unlikely to represent the asymmetry of the explosion, as interaction is expected to have played a part in shaping the ejecta of SN 2009ip. \citet{Mauer2014} compared the polarimetric properties of SN 2009ip to those of Type IIb SN 1993J, noticing that $\sim$50 days after peak the SNe are similar in degree of polarisation and polarisation angle, but that around peak SN 1993J displayed more variation around the average $p$ and $\theta$ than SN 2009ip.

If there is an underlying Type IIP SN in SN 2009ip we could expect to see a similar increase in the continuum polarisation at $\sim$100 days after explosion \citep{Arc2012}. This corresponds to the geometric discontinuity observed at the end of the plateau phase in other Type IIP SNe \citep{Leo2006, Chor2010, Mau2007}. Assuming that the onset of the 2012a event represents the SN explosion epoch, the beginning of our observations corresponds to $\sim$100 days. There are no indications, however, of a discontinuity in the geometry at this time, as the continuum polarisation appears only to decrease between the peak of 2012b \citep{Mauer2014} and the beginning of these observations. Should the onset of 2012b be taken as the explosion epoch the plateau phase would end around 1$^{\mathrm{st}}$ January 2013, after SN 2009ip had set and our observations finished. 

Spectropolarimetric observations have been performed for only a few Type IIn SNe, namely SN 1997eg, SN 1998S and SN 2010jl \citep{Hoffman2008, Leo2000, Wan2001,Pat2011, Bau2012}. Similarly to SN 2009ip, the other Type IIn SNe exhibited significant continuum polarisation. SNe 1997eg and 2010jl were both polarised to $p\sim$2\% \citep{Hoffman2008, Pat2011, Bau2012}, for SN 1998S the continuum polarisation increased from $\sim$2\% to $p\sim$3\% upon removal of the ISP \citep{Leo2000}. Under a different estimate for the ISP, \citet{Wan2001} found continuum polarisation for SN 1998S in the range of 1.6-3\%. The maximum continuum polarisation observed in the data presented here is $\sim$0.8\% at $+$35 days and $\sim$1.7\% at around the 2012b peak as observed by \citet{Mauer2014}; implying a lower degree of asymmetry for SN 2009ip compared to the other Type IIn SNe. Given the uncertainty in where the continuum polarisation is generated and what governs the geometry of the ejecta in Type IIn SNe, it is difficult to determine if this means a greater asymmetry in the explosion mechanism or in the CSM around the other SNe. For both SN 1998S and 1997eg the continuum polarisation is thought to be produced in the ejecta, as is the case for SN 2009ip after $+$35 days. This could imply a greater asymmetry in the explosion mechanism for those SNe in comparison to SN 2009ip. 

For SN 2009ip we observe no significant increase in the degree of polarisation in the broad emission wings of \halpha\,\,compared to the core of the broad component. This is not the case for SN 2010jl, where the degree in polarisation is enhanced in the wings and interpreted as due to up-scattering of the photons in a dense fully ionised CSM \citep{Pat2011}. The lack of more highly polarised wings in SN 2009ip indicates that there is little reprocessing of the photons, at least from $+$35 days, and implies a line-forming region in the fast-moving ejecta, similarly to the broad emission line-forming regions in SNe 1997eg and 1998S \citep{Hoffman2008, Leo2000}.

The similar behaviour of the broad H and He\,{\sc i} $\lambda 5876$ lines in SN 2009ip suggests that the ejecta are both H and He rich, the polarisation at low velocity for both of these features indicates that the surrounding medium is also H and He rich. This also seems to be the case for SN 2010jl, where helium appears to be polarised in the same manner as the broad hydrogen features, suggesting that the CSM surrounding SN 2010jl is also helium rich \citep{Pat2011, Bau2012}. \citeauthor{Hoffman2008} found that for SN 1997eg, He\,{\sc i} $\lambda5876$ was polarised via electron-scattering in the ellipsoidal ejecta in which the line formed. The line also behaved differently compared to the Balmer lines on the $q$-$u$ plane. They concluded that the ejecta for SN 1997eg were helium rich while the CSM was composed of hydrogen. This could imply a difference in mass loss history between the progenitors of SN 1997eg and SNe 2010jl and 2009ip. For SN 1998S, the mass loss history was determined to be episodic \citep{Leo2000} as has been observed with SN 2009ip.

For SNe 1997eg, 1998S and 2010jl, the geometry of the CSM has been inferred to be disk-like \citep{Hoffman2008, Leo2000, Wan2001,Pat2011, Bau2012, And2011, Cha2015}.  In addition, the spectra of the more recent SN 2015bh, a SN 2009ip-like event, have also displayed evidence of asymmetry in the ejecta and CSM, with a disk-like geometry proposed \citep{Eli2016, Tho2016, Gor2016}. The inferred disk or torus-like CSM surrounding SN 2009ip is consistent with the geometries suggested for other Type IIn SNe. In general, the inferred geometrical configurations for each of the Type IIn SNe are similar (aspherical ejecta surrounded by a disk-like or torus CSM) however differences remain in the composition of the CSM and how it induces a polarised signal. For SN 1997eg and SN 2010jl, electron-scattering in the CSM is responsible for the observed polarisation, however for SN 2009ip the equatorial disk obscures the photosphere inducing the polarised signal. The evolution of the polarimetry corroborates what other observables such as the double-peaked light curve, the observed broad lines and the sustained eruptive mass loss demonstrated: SN 2009ip is not quite the same as a typical Type IIn SN.


\section{Conclusions}
\label{sec:09ip_conclusions}

During observations covering the period from $+$35 to $+83$ days (with respect to the UV maximum of the 2012b re-brightening) SN 2009ip exhibited significant continuum and line polarisation. The continuum was intrinsically polarised at the 0.3-0.8\% level throughout the observations, implying asphericities of $\sim$10-15\% in the shape of the photosphere. The observed degree and angle of the continuum polarisation were shown to depend on the choice of the ISP, and therefore interpretation of these results should be approached with caution.

Significant polarisation was found associated with the absorption profiles \halpha, \hbeta, He\,{\sc i} $\lambda5876$ and Ca\,{\sc ii} IR triplet. Mirroring the total flux profile, the polarised flux of \halpha\,\,was composed of multiple components. Both broad and narrow depolarisation was observed to be associated with narrow and broad P Cygni emission profiles. Narrow peaks in the polarisation ($\sim$1\%) were associated with absorption features at low ($\sim -$1000 \kms) and high velocities. In contrast to the behaviour of the high-velocity material, the polarisation associated with the low-velocity narrow \halpha\,\,P Cygni absorption exhibited no temporal evolution. The distinct polarisation angles and differing behaviour revealed the independent geometries of the fast-moving ejecta and the highly-aspherical, slow-moving line-forming region generated in the CSM. The similarity of polarisation peaks for \halpha, \hbeta\,\,and He\,{\sc i} $\lambda5876$ suggested a shared structure for hydrogen and helium at both high ($\sim$$-$13,800 \kms) and low velocities.  

A Monte Carlo simulation of the photosphere and line-forming region of the low-velocity \halpha\,\,feature was performed in an effort to reproduce the geometry of the CSM. A disk-like structure observed edge-on or inclined by 14$\pm$2$^{\circ}$ to the line of sight successfully replicated the observed Stokes parameters at $+$35 and $+$68 days. The inclined-disk model was expected to begin to cover the photosphere, producing an absorption feature, just prior to the 2012b maximum and disappear at $\sim+$70 days. This was in agreement with the observed disappearance of the absorption line between $+$68 and $+$83 days. The edge-on disk model was difficult to reconcile with the absence of a low-velocity P Cygni absorption during the 2012a outburst. In addition, the observation of a thin disk-like structure edge-on would require specific alignment.
   
The inclined disk-like CSM was thus favoured. In this scenario, the low-velocity \halpha\,\,polarisation is produced by the same material that interacts with the ejecta at earlier times to power the 2012b maximum. We identified two distinct phases in the evolution of SN 2009ip. In the first phase at $\sim+$35 days, the disk structure induces a strong polarisation by blocking the ejecta photosphere that is elongated in the plane perpendicular to the CSM. Two scenarios naturally explain the orthogonality of the ejecta and the CSM. In the case of a pre-existing disk (prior to the 2012a outburst), the ejecta will more easily expand in the orthogonal direction. Alternatively, an aspherical explosion in the 2012a outburst expanding into a spherical CSM will result in an oblate or disk-like structure in the perpendicular direction. The observations made in December 2012 ($\sim+$70 days) represent a new phase in the polarimetric data where the inner core of the explosion is revealed upon recession of the photosphere. At this stage the photosphere rotates to align with the CSM.

The inferred geometry of SN 2009ip fits in quite well with the observed bipolar nebulae surrounding Luminous Blue Variables. The possibility of pre-existing bipolar lobes similar to the Homunculus nebula of $\eta$ Car is not ruled out, however the polarimetric observations here are not suggestive of the existence of such lobes. The geometrical configuration of the CSM surrounding SN 2009ip is similar to that inferred for other Type IIn SNe, however differences remain in the composition of the previously ejected material. The CSM here is both hydrogen and helium rich in contrast to the hydrogen rich CSM observed around other SNe. The evolution of the narrow and broad line profiles imply that SN 2009ip is not quite a typical Type IIn.

The observation of the disk geometry for the CSM may imply the terminal explosion of the LBV progenitor in one of the 2012 outbursts, however, this depends on the origin of this geometry. The implication of terminal SN explosion energies is only valid if the disk-like CSM pre-dates the 2012a explosion. A highly bipolar explosion colliding with a spherical circumstellar shell may result in the formation of this type of structure during the 2012b event, in which case the ejecta would interact with a large fractional area of the CSM. The 2012b light curve could then be powered through interaction with a non-terminal eruption. It is not clear that the inferred orientation of the inclined-disk here is consistent with that of \citet{Mauer2014}; simulations of the interaction of the ejecta with both the disk-like CSM and an originally spherical CSM are required in order to investigate if the continuum polarisation observed at the peak of the interaction can be replicated.

\section{Acknowledgements}
We thank the ESO Director General for awarding discretionary time for the observations and the observers at Paranal for acquiring these high-quality observations. ER was supported by a PhD studentship awarded by the Department of Education and Learning of Northern Ireland. The research of JRM is supported through a Royal Society University Research Fellowship. ER thanks Morgan Fraser, Cosimo Inserra and Anders Jerkstand for useful discussions.

\bibliographystyle{mnras}

\appendix
\section{Line Polarisation of SN 2009ip}

\begin{table*}
\label{table:linepol}
\caption{SN 2009ip line polarization.}
 \begin{tabular}[alignment]{c c c c c c c c c c}
Phase (days) $\dag$ & Species & $\lambda$$_{0}$ (\ang) & $\lambda$ (\ang) & v (\kms) & $p_{\rm line}$ (\%) & $\sigma p_{\rm line}$ (\%) & $\theta_{\rm line}$ ($^{\circ}$) & $\sigma\theta_{\rm line}$  ($^{\circ}$) & Grism \\
\hline
 \hline
& & & & Broad features & & & & &\\ 
 \hline
 \\
35 & \halpha & 6563 & 6300-6500 & $-$12000 - $-$3000 & 0.34 & 0.09 & 128 & 9 &1200R\\
42 & \halpha & 6563 & 6300-6500 &  $-$12000 - $-$3000 & 0.43 & 0.13 & 125 & 18 & 1200R\\
68 & \halpha & 6563 &6300-6500 &  $-$12000 - $-$3000 & 0.65 & 0.19 & 85 & 42 & 1200R\\
\\
 \hline
 & & & & Low velocity features & & & & &\\ 
 \hline
 \\
35 & \halpha & 6563 & 6538 & $-$1150 & 0.95 & 0.28 & 34 & 9 &1200R\\
35 & \halpha & 6563 & 6533 & $-$1400 & 0.30 & 0.23 & 50 & 15 & 300V\\ 
35 & \hbeta & 4861 & 4853 & $-$500 & 0.4 & 0.3 & 44 & 24 & 300V\\ 
35 & He\,{\sc i} & 5876 & 5853 & $-$1160 & 1.4 & 0.6 & 32 & 16 & 1200R \\
\\
42 & \halpha & 6563 & 6543 & $-$900 & 1.0 & 0.4 & 50 & 11 &1200R\\ 
42 & \hbeta & 4861 & 4838 & $-$1400 & 1.1 & 0.3 & 98 & 9 & 300V\\ 
\\
68 & \halpha & 6563 & 6538 & $-$1150 & 0.9 & 0.7 & 47 & 21 &1200R\\ 
\\
 \hline
 & & & & Higher velocity features & & & & &\\ 
 \hline
 \\
35 & \hbeta & 4861 & 4643 & $-$14100 & 0.5 & 0.3 & 48 & 20 & 300V\\
35 & He\,{\sc i} & 5876 & 5618 & $-$13800 & 0.7 & 0.3 & 62 & 12 & 300V \\
\\

42 & He\,{\sc i} & 5876 & 5588 & $-$15500 & 0.7 & 0.4 & 61 & 15 & 300V\\
42 & \halpha & 6563 & 6408 & $-$7300 & 1.0 & 0.4 & 141 & 12 &1200R\\
\\
 
64 & \halpha & 6563 & 6307 & $-$12150 & 0.6 & 0.4 & 109 & 18 & 300V\\ 
64 & \halpha & 6563 & 6352 & $-$9900 & 0.7 & 0.3 & 121 & 14 & 300V\\ 
64 & He\,{\sc i} & 5876 & 5692 & $-$9700 & 0.7 & 0.3 & 124 & 13 & 300V\\
64 & \halpha & 6563 & 6412 & $-$7000 & 0.8 & 0.3 & 139 & 12 & 300V\\
64 & Ca\,{\sc ii}& 8542 & 8408 & $-$4700 & 1.1 & 0.4 & 111 & 9 & 300V\\ 
64 & He\,{\sc i} & 5876 & 5827 & $-$2500 & 1.0 & 0.3 & 127 & 10 & 300V\\
64 & Ca\,{\sc ii} & 8662 & 8573 & $-$3100 & 0.8 & 0.4 & 95 & 14 & 300V\\ 
\\

73 & \hbeta & 4861 & 4625 & $-$15250 & 1.4 & 0.9 & 62 & 19 & 300V\\
73 & He\,{\sc i} & 5876 & 5705 & $-$8900 & 1.2 & 0.9 & 104 & 22 & 300V\\ 
73 & \halpha & 6563 & 6425 & $-$6400 & 1.1 & 0.9 & 128 & 23 & 300V\\ 
73 & \hbeta & 4861 & 4760 & $-$6300 & 1.5 & 0.9 & 114 & 17 & 300V\\
73 & Ca\,{\sc ii} & 8542 & 8375 & $-$5900 & 3.4 & 0.9 & 139 & 9 & 300V\\
73 & Ca\,{\sc ii} & 8662 & 8570 & $-$2100 & 1.7 & 0.9 & 118 & 14 & 300V\\
\hline
 \end{tabular}
  \vspace{2mm}
    \begin{flushleft}
  \dag Relative to the UV maximum of the 2012b re-brightening on the 3$^{\rm rd}$ October 2012\\
  \end{flushleft}
  \label{linetable}
\end{table*}

\section{Simulated polarisation of the low-velocity \halpha\,\,feature}
\newgeometry{margin=1.5cm} 
\begin{landscape}

\begin{table}
\label{table:model}
\caption{Origin of the low-velocity \halpha\,\,polarisation: Simulation results}
 \begin{tabular}[alignment]{c c c c c c c c c c c c c c c}

Phase \dag & $q_{\rm cont}$ (\%) & $u_{\rm cont}$ (\%) & $p_{\rm cont}$ (\%) & $\theta_{\rm cont}$ ($^{\circ}$) & Wavelength (\ang) & $q_{\rm obs}$ (\%)$\ddag$ & $u_{\rm obs}$ (\%) & $p_{\rm obs}$ (\%)& $\theta_{\rm obs}$ ($^{\circ}$) & $q_{\rm line}$ (\%)& $u_{\rm line}$ (\%)& $p_{\rm line}$  (\%)& $\theta_{\rm line}$ ($^{\circ}$) & Absorption Depth (\%)\\
\hline
\hline
& & & & & Observed data & & & & & & & & &\\
\hline
$+$35 & 0.09$\pm$0.20 & 0.74$\pm$0.19 & 0.74$\pm$0.20 & 41$\pm$8 & 6538 & 0.42$\pm$0.33 & 1.68$\pm$0.29 & 1.74$\pm$0.32 & 38$\pm$5 & 0.33$\pm$0.28 & 0.94$\pm$0.22 & 0.95$\pm$0.28 & 34$\pm$9 & 6\\
$+$68 & -0.17$\pm$0.52 & -0.3$\pm$0.6 & 0.6$\pm$0.4 & 119$\pm$23 & 6538 & -0.34$\pm$0.23 & 0.97$\pm$0.16 & 1.00$\pm$0.48 & 53$\pm$14 & -0.2$\pm$0.6 & 1.2$\pm$0.6 & 1.0$\pm$0.7 & 47$\pm$21 & \\
\\
\hline
& & & & & Edge-on disk-like CSM ($z$ = 11 {\sc au}) & & & & & & & & &\\
\hline
$+$35 & 0.05 & 0.41 & 0.41 & 41 & 6538 & 0.07 & 1.43 & 1.43 & 44 & 0.02 & 1.02 & 1.02 & 45 & 16\\
$+$68 & -0.26 & -0.5 & 0.56 & 121 & 6538 & -0.21 & 0.40 & 0.45 & 59 & 0.05 & 0.90 & 0.90 & 43 & 15\\
\\
\hline
& & & & & Inclined disk-like CSM ($i$ = 14$^{\circ}$) & & & & & & & & &\\
\hline
$+$35 & 0.04 & 0.66 & 0.66 & 43 & 6538 & 0.41 & 1.69 & 1.69 & 44 & -0.002 & 1.03 & 1.03 & 45 & 9\\
$+$68 & -0.25 & -0.45 & 0.51 & 121 & 6538 & -0.27 & 0.36 & 0.45 & 63 & -0.02 & 0.81 & 0.81 & 45 & 6\\
\hline
 \end{tabular}
   \vspace{2mm}
    \begin{flushleft}
  \dag Relative to the UV maximum of the 2012b re-brightening on the 3$^{\rm rd}$ October 2012\\
  \ddag In order to correct for the depolarisation across \halpha, $q_{\rm obs}$ and $u_{\rm obs}$ were determined from the vector addition of the continuum and line polarisation measurements listed above. The uncertainties were propagated and $p_{\rm obs}$ and $\theta_{\rm obs}$ calculated.\\
  \end{flushleft}

\end{table}
\end{landscape}
\restoregeometry

\end{document}